\documentclass[aps,prd,amsmath,floats,floatfix,twocolumn,superscriptaddress,
  nofootinbib,showpacs]{revtex4} 
\usepackage{graphicx}
\usepackage{bm}
\usepackage{color}

\newcommand {\beq}{\begin{equation}}
\newcommand {\eeq}{\end{equation}}
\newcommand {\bea}{\begin{eqnarray}}
\newcommand {\eea}{\end{eqnarray}}

\newcommand{\Eadm}{M_{\text{ADM}}}
\newcommand{\Jadm}{J_{\text{ADM}}}

\newcommand{\Cornell}{\affiliation{Center for Radiophysics and Space
    Research, Cornell University, Ithaca, New York, 14853, USA}}
\newcommand{\Caltech}{\affiliation{Theoretical Astrophysics 130-33,
    California Institute of Technology, Pasadena, California 91125, USA}}

\begin{document}
\vspace{-2.5cm} 

\title{Initial data for black hole--neutron star binaries: a flexible, high-accuracy spectral method.}

\author{Francois Foucart} \Cornell
\author{Lawrence E. Kidder} \Cornell
\author{Harald P. Pfeiffer} \Caltech
\author{Saul A. Teukolsky} \Cornell

\date{\today}

\begin{abstract}
We present a new numerical scheme to solve the initial value problem for black hole--neutron star binaries. 
This method takes advantage of the flexibility and fast convergence of a multidomain spectral representation of the initial 
data to construct high-accuracy solutions at a relatively low computational cost.
We provide convergence tests of the method for both isolated neutron stars and irrotational binaries. In the second case, we show 
that we can resolve the small inconsistencies that are part of the quasiequilibrium formulation, and that these 
inconsistencies are significantly smaller than observed in previous works.
The possibility of generating a wide variety of initial data is also demonstrated through two new configurations inspired by results from binary black holes. 
First, we show that choosing a modified Kerr-Schild conformal metric instead of a flat conformal metric allows for the construction of quasiequilibrium binaries 
with a spinning black hole. Second, we construct binaries in low-eccentricity orbits, which are a better approximation to astrophysical binaries than
quasiequilibrium systems.
\end{abstract}

\maketitle

\section{Introduction}

Over the last few years, the prospect of gravitational wave detection by ground based experiments such as LIGO ~\cite{LIGO}
and VIRGO ~\cite{VIRGO} has encouraged rapid developments in the field of numerical relativity. Most of that effort was
aimed at the evolution of compact binaries, sources of waves potentially observable by those detectors.
Binary neutron stars were the first to be successfully evolved in a fully relativistic framework, and have been studied
regularly over the last eight years ~\cite{Shibata2000,Duez2003,Miller2004,Jin2007,Shibata2005,Anderson2008,Baiotti2008}. 
Evolutions of binary black holes (BBH) followed a few years later ~\cite{Pretorius2005,Baker2006,Campanelli2006}, and 
continue to be an extraordinarily active area of research (see ~\cite{Pretorius2008} and references therein).

The third type of compact binary, black hole--neutron star (BH-NS) binaries, has not been as widely studied yet. The
evolution of the black hole singularity and the presence of matter combine the difficulties of evolving both binary black
holes and binary neutron stars. And the system has its own specific challenges, notably the accretion of the neutron star
matter onto the black hole. Such binaries are, however, worth studying not only for their interest as gravitational
wave sources, but also as potential sources of gamma-ray bursts ~\cite{Lee2007}. The first evolutions of such systems
were announced very recently ~\cite{Etienne2008,Shibata2007}, and such evolutions will allow more extensive
study of their wave emission, merger, accretion disk formation, and so on. 

The choice of a suitable initial configuration for binary evolutions has been a long-standing problem. Not only do the
Einstein equations include constraints on the initial data, but also choosing a starting point that represents a
realistic astrophysical situation is not trivial. Because of their computational cost, numerical simulations of compact binaries 
usually start just a few orbits away from merger. The two objects are close enough that the nonlinearity of the Einstein equations is important. 
In that regime, there is no known way of prescribing the exact state of
the system. The most common assumption is that the binary has had time to settle into a quasiequilibrium state, the system being approximately 
time-independent in the corotating frame. Furthermore, as the viscous forces within the star are expected to be small,
we do not expect much change in the spin of the star as the orbital radius decreases. 
For an initially nonspinning neutron star, this would lead to an irrotational
velocity profile, another standard assumption. Because of gravitational wave emission, there is no exact equilibrium state, however.
Accordingly, these conditions cannot be perfectly satisfied, a problem we will
discuss in more detail later on.

Previous results on initial data for BH-NS evolutions include the early work of Taniguchi et al. ~\cite{Taniguchi2005} and
Sopuerta et al. ~\cite{Sopuerta2006}, as well as more recent initial configurations generated by Taniguchi
et al.~\cite{Taniguchi2006,Taniguchi2007,Taniguchi2008} and Grandclement ~\cite{Grandclement2006}. Both
Taniguchi and Grandclement use codes based on the {\sc lorene} package ~\cite{LORENE}, and their most recent publications are 
similar in accuracy, computational cost, and numerical results.

In this paper, we present an alternative numerical scheme for the solution of this problem. Our code is based on the 
spectral
elliptic solver ({\sc spells}) developed by the Cornell-Caltech collaboration ~\cite{Pfeiffer2003b}, and originally used by Pfeiffer 
~\cite{Pfeiffer2002,Pfeiffer2003} for the study of binary black holes (BBH) initial data. For our numerical tests, the mathematical formulation
of the problem will be very similar to ~\cite{Taniguchi2008} and ~\cite{Grandclement2006}, allowing
easy evaluation of the performance of our code. 

Our motivation for using {\sc spells} is the remarkable flexibility of its multidomain spectral methods. This allows us to
efficiently adapt the configuration of our numerical grid to the geometry of the system and yields high-precision
results at a very reasonable computational cost.
As we will see in Sec. \ref{sec:formalism}, elliptic equations form the core of the initial data problem. Using 
{\sc lorene},  each of those equations has to be approximated by two Poisson equations, with coupled source terms.
These two sets of equations are then solved through an iterative method.
The variables are fields with an
approximate spherical symmetry around one of the compact objects. However, as the source terms for the BH fields include terms 
centered around the NS, obtaining high-precision initial data requires a large angular resolution.

With {\sc spells}, by contrast we do not have to limit ourselves to spheres around the compact objects. We can instead choose among
a wide variety of subdomain geometries and coordinate mappings. As the basis functions of our spectral expansion are
more adapted to the geometry of the solution, a significantly smaller number of collocation points are necessary
to reach a given accuracy.

In Sec. \ref{sec:ConvBin}, we will see that the main sources of error in our initial data are the approximations
introduced by the quasiequilibrium formulation. Using {\sc spells}, we can rapidly solve the initial data problem for a large
variety of configurations to a precision allowing us to resolve these errors. We will show that they appear
to be significantly lower than quoted in ~\cite{Taniguchi2008,Grandclement2006}. For the closest binaries, when the distortion of the
star limits the precision of any spectral method, such precision is no longer possible --- at least using our current numerical
grid. But our error remains reasonable, reaching the level of the deviations from equilibrium mentioned in 
~\cite{Taniguchi2008} for the most extreme cases.

In addition to the high-precision initial data our results provide for evolutions of BH-NS binaries, they should also make it 
possible to explore the limits of the quasiequilibrium formalism. Such studies are already possible for BBH binaries, as shown in ~\cite{Cook2004}.
On the BH horizon, the deviations from equilibrium computed in ~\cite{Cook2004} are similar to our own results. 

Using {\sc spells}, we are also able to study initial data for a spinning BH by abandoning the assumption of conformal flatness. Earlier results showed
that initial configurations built using a Kerr-Schild conformal metric were significantly inferior to their conformally flat counterparts 
~\cite{Taniguchi2006,Taniguchi2007}. Here, adapting a method developed by Lovelace for BBH ~\cite{Lovelace2007}, we show that a \emph{modified}
Kerr-Schild metric can lead to high-precision initial data. In Sec.\ \ref{sec:SpinBH} we
present our results for spinning and nonspinning black holes using this modified Kerr-Schild conformal metric. 

We review the formulation of the initial value problem in Sec.\ \ref{sec:formalism}, and
present in more detail our numerical methods in Sec.\ \ref{sec:NumMeth}. Then, in Sec.\ \ref{sec:NumTests}, we discuss some tests of our code, 
including isolated stars and binaries that are directly comparable to previous results. Through 
convergence tests, we obtain a good estimate of the amplitude of constraint violations and of our error in global quantities
such as the ADM (Arnowitt-Deser-Misner) energy and linear and angular momentum. Such convergence tests for fully consistent initial data in 
the presence of matter have, to our knowledge, only 
been published previously in the case of NS-NS binaries (see for example ~\cite{Gourgoulhon2001}, 
specifically Figs.\ 4 to 7), and up to relative precisions slightly better than $10^{-5}$. Our estimates will confirm
that we are able to resolve deviations from quasiequilibrium except for strongly distorted stars.

Finally, adapting a method developed by Pfeiffer et al. ~\cite{Pfeiffer2008} for BBH binaries, we demonstrate the possibility 
of reducing the eccentricity of the system, leading to initial configurations more realistic than 
quasiequilibrium orbits.

\section{The initial data problem}
\label{sec:formalism}

The construction of initial data on a spatial slice containing matter typically involves two types of conditions.
First, from the Einstein equations we know that any initial data will have to satisfy the Hamiltonian and momentum constraints, 
which we will write as a set of elliptic equations. Second, we want the resulting 
configuration to represent a physically reasonable
situation. The mass of each object, its spin, their initial separation, and the ellipticity of the orbit are all parameters
we want to control, and the initial state and physical properties of the fluid have to be carefully chosen.
In this section, we
will describe the different equations used to enforce those conditions, and their formulation 
in our numerical solver.

\subsection{Constraints}
\label{sec:Constraints}
We impose the constraints on our initial spatial slice by solving the extended conformal thin sandwich (XCTS)
system, a set
of 5 elliptic equations based on the conformal thin sandwich decomposition proposed by York ~\cite{York1999}.
Here, we start from the formulation used by Pfeiffer~\cite{Pfeiffer2003} for BBH binaries, adding the matter contribution 
as fixed source terms in the XCTS equations. 

The metric tensor is written in its 3+1 form:
\bea
\nonumber
ds^2&=&g_{\mu \nu}dx^{\mu}dx^{\nu}\\
&=&-\alpha^2dt^2+\gamma_{ij}\left(dx^i+\beta^idt\right)\left(dx^j+\beta^jdt\right),
\eea
where $\alpha$ is the lapse, $\beta^i$ the shift, and $\gamma_{ij}$ the 3-metric induced on a spatial
slice at constant $t$. The normal ${\bf n}$ to such a slice and the tangent to the coordinate line ${\bf t}$ are 
then related by
\beq
t^{\mu}=\alpha n^{\mu}+\beta^{\mu}.
\eeq
We treat the matter as a perfect fluid and write the stress-energy tensor as
\beq
T_{\mu \nu}=\left(\rho + P\right) u_{\mu}u_{\nu}+P g_{\mu \nu},
\eeq
where $\rho$ is the fluid energy density, $P$ its pressure, and $u_{\mu}$ its 4-velocity.
In practice, we will use projections of $T_{\mu\nu}$:
\bea
E&=&T^{\mu\nu}n_{\mu}n_{\nu}=(\rho + P)\frac{1}{1-\gamma_{ij}U^iU^j} -P,\\
S&=&\gamma^{ij}\gamma_{i\mu}\gamma_{j\nu}T^{\mu\nu}=E+3P-\rho,\\
J^i&=&-\gamma^{i}_{\ \nu}T^{\nu\tau}n_{\tau}=U^i\frac{1}{1-\gamma_{ij}U^iU^j}(\rho+P),
\eea
where $U^i$ is the fluid 3-velocity in the inertial frame, defined in terms of the 4-velocity ${\bf u}$,
the normal ${\bf n}$ to the spatial slice studied, and the Lorentz factor $\gamma_n$ as
\beq
{\bf u}=\gamma_n({\bf n}+{\bf U}).
\eeq
If the system is close to equilibrium, it
is convenient to choose the coordinate system so that $\partial_t$ is an approximate Killing vector. We will
thus try to solve the system in coordinates comoving with the binary.
In such a coordinate system, the shift increases in magnitude with the distance from the center of rotation and
diverges at spatial infinity.
This is a difficulty for numerical solvers. Furthermore, to control the eccentricity of the binary, we choose
to give the system an initial radial velocity of the form ${\bf v} = \dot{a}_0 {\bf r}$. 
This also leads to a diverging term in the shift at large distances.

We thus further decompose the shift vector as
\beq
\label{eq:shiftdec}
{\bm\beta}={\bm\beta}_0+{\bm\Omega} \times {\bm r}+\dot{a}_0 {\bm r},
\eeq
where ${\bm\beta}_0$ is the shift in the inertial frame and ${\bm\Omega}$ the orbital angular velocity of the system. 
In practice, we solve for ${\bm\beta}_0$ instead of ${\bm\beta}$, as ${\bm\beta}_0$ conveniently vanishes at spatial infinity.
We turn now to the extrinsic curvature, defined as
\beq
\label{eq:Kdef}
K_{\mu \nu}=-\tfrac{1}{2} {\cal L}_n g_{\mu \nu},
\eeq
where ${\cal L}_n$ is the Lie derivative along the normal ${\bm n}$.
In the conformal thin sandwich formalism, $K_{\mu \nu}$ is divided into its trace $K$ and trace-free part $A^{ij}$:
\beq
\label{eq:Kdec}
K^{ij}=A^{ij}+\tfrac{1}{3}\gamma^{ij}K.
\eeq
The decomposition is completed by the use of conformal transformations according to the scheme
\footnote{A conformal transformation of the matter quantities $E$, $S$ and $J^i$ is necessary for the Hamiltonian constraint 
to have a unique solution ~\cite{York1979}. But different choices for the ratio between
conformal and physical quantities are valid. Our choice of $\phi^{6}$, which differs from ~\cite{Pfeiffer2003},
guarantees that volume integrals of the matter terms for fixed $\tilde E$, $\tilde S$ and $\tilde J^i$ are
independent of the conformal factor $\phi$. Indeed, the physical volume element on the spatial slice is 
$dV=\phi^6 \sqrt{\tilde \gamma} d^3x$, where  $\tilde \gamma$ is the determinant of the conformal metric, and thus
$\int{EdV}=\int{\tilde E d^3x}$. The full XCTS system is known to have non-unique solutions for vacuum \cite{PfeifferYork2005,Lovelace2008}; 
this may carry over to space-times with matter, but we have not observed non-uniqueness in the course of the present work.}:
\bea
\gamma_{ij}&=&\phi^4 \tilde \gamma_{ij},\\
E&=&\phi^{-6} \tilde E,\\
S&=&\phi^{-6} \tilde S,\\
J^i&=&\phi^{-6} \tilde J^i,\\
A^{ij}&=&\phi^{-10}\tilde A^{ij},\\
\alpha&=&\phi^6 \tilde \alpha.
\eea

Denoting the time derivative of the conformal spatial metric by $\tilde u_{ij}=\partial_t\tilde g_{ij}$, 
Eqs. (\ref{eq:Kdef}) and (\ref{eq:Kdec}) link $\tilde A^{ij}$ and the shift by
\beq
\tilde A^{ij}=\frac{1}{2\tilde \alpha}\big[(\tilde L \beta)^{ij}-\tilde u^{ij}\big],
\eeq
where the conformal longitudinal operator $\tilde L$ is
\beq
(\tilde LV)^{ij}=\tilde \nabla^i V^j+\tilde \nabla^j V^i - \frac{2}{3}\tilde \gamma^{ij}\tilde \nabla_kV^k.
\eeq
The XCTS formulation of the constraints is then a set of 5 coupled elliptic equations, with the conformal factor $\phi$, 
the densitized lapse $\alpha \phi=\tilde \alpha \phi^7$, and the shift ${\bm \beta}$ (or, in practice, the inertial shift ${\bm \beta}_0$) as variables:
\bea
\label{eq:XCTSShift}
2\tilde \alpha\bigg\{\tilde \nabla_j \big[\frac{1}{2\tilde \alpha}(\tilde L \beta)^{ij}\big]-\tilde \nabla_j \bigg(\frac{1}{2\tilde \alpha}\tilde u^{ij}\bigg)&& \\ \nonumber
  -\frac{2}{3}\phi^6\tilde \nabla^i K -8\pi \phi^4 \tilde J^i\bigg\}&=&0
\eea
\bea
\label{eq:XCTSPhi}
\tilde \nabla^2 \phi -\frac{1}{8}\phi \tilde R - \frac{1}{12}\phi^5 K^2 + \frac{1}{8}\phi^{-7}\tilde A_{ij} \tilde A^{ij} &&\\ \nonumber
+2\pi \phi^{-1} \tilde E&=&0
\eea
\bea
\label{eq:XCTSNphi}
\tilde \nabla^2 \left(\tilde \alpha \phi^7 \right) - \left(\tilde \alpha \phi^7 \right)\bigg[\frac{1}{8}\tilde R+\frac{5}{12}\phi^4K^2+\frac{7}{8}\phi^{-8}\tilde A_{ij}\tilde A^{ij} && \\ \nonumber
 +2\pi \phi^{-2}\left(\tilde E+2\tilde S\right) \bigg]=-\phi^5\left(\partial_t K - \beta^k\partial_k K\right).&&
\eea
Here, $\tilde E$, $\tilde S$, and $\tilde J^i$ determine the matter content of the slice, and we are free
to choose $\tilde \gamma_{ij}$, $\tilde u_{ij}$, $K$, and $\partial_t K$. Eqs. (\ref{eq:XCTSShift}) and 
(\ref{eq:XCTSPhi}) are the momentum and Hamiltonian constraints, while Eq. (\ref{eq:XCTSNphi}) can be
derived from the evolution equation for $K^{ij}$. (For more details on the XCTS system, and its derivation, see
~\cite{PfeifferYork2003}.)

For quasiequilibrium initial conditions, a natural choice for the free variables is to set the time 
derivatives to zero. The choice of $\tilde \gamma_{ij}$ and $K$ is, however, less obvious. Taniguchi et al. 
~\cite{Taniguchi2006,Taniguchi2007} showed that a conformally flat metric ($\tilde \gamma_{ij}=\delta_{ij}$) with maximal 
slicing ($K=0$) gives good results --- better than using a Kerr-Schild background at least. For the tests
in this paper, we will make the same choice. In Sec.\ \ref{sec:SpinBH}, however, we will show that different
choices lead to acceptable initial data, and make it possible to construct spinning BHs. 

\subsection{Hydrostatic equilibrium}
\label{sec:hydroeq}
The initial state of the matter within the neutron star is, in general, unknown. However, we can
make some reasonable approximations. First, we will require the fluid to be in a state of
hydrostatic equilibrium in the comoving frame. Following the method described by Gourgoulhon et al.
~\cite{Gourgoulhon2001}, we use the first integral of the Euler equation,
\beq
\label{eq:EulerConstant}
h\alpha \frac{\gamma}{\gamma_0}={\rm constant},
\eeq
where $h$ is the fluid enthalpy and we define the Lorentz factors
\bea
\gamma&=&\gamma_n\gamma_0\left(1-\gamma_{ij}U^iU^j_0\right),\\
\label{eq:Lorentz}
\gamma_0&=&\left(1-\gamma_{ij}U^i_0U^j_0\right)^{-1/2},\\
\gamma_n&=&\left(1-\gamma_{ij}U^iU^j\right)^{-1/2},\\
\label{eq:comvel}
U^i_0&=&\frac{\beta^i}{\alpha}.
\eea
As before, $U^i$ is the fluid 3-velocity in the inertial frame, while $U^i_0$ is the 3-velocity of a comoving observer. For a corotating binary, we simply have $U^i=U^i_0$, while
for an irrotational configuration, there should exist a velocity potential $\Psi$ ~\cite{Gourgoulhon2001} such that
\beq
U^i=\frac{\phi^{-4}\tilde\gamma^{ij}}{h\gamma_n}\partial_j\Psi .
\eeq
The equation of continuity is then
\beq
\label{eq:continuity}
\frac{\rho_0}{h}\nabla^{\mu}\nabla_{\mu}\Psi+\left(\nabla^{\mu}\Psi\right)\nabla_{\mu}\frac{\rho_0}{h}=0,
\eeq
where $\rho_0$ is the baryon density. This is an elliptic equation in $\Psi$, which we can rewrite more explicitly in our variables as
\bea
\label{eq:HydroPot}
&& \rho_0\bigg\{-\tilde\gamma^{ij}\partial_i\partial_j\Psi+\bigg[\tilde\gamma^{ij}\tilde\Gamma^k_{ij}+\tilde\gamma^{ik}\partial_i\left(\ln{\frac{h}{\alpha \phi^2}}\right)\bigg]\partial_k\Psi 
 \\ \nonumber
&& +\frac{h\beta^i\phi^4}{\alpha}\partial_i\gamma_n+hK\gamma_n\phi^4\bigg\} =\tilde\gamma^{ij}\partial_i\Psi\partial_j\rho_0
-\frac{h\gamma_n\beta^i\phi^4}{\alpha}\partial_i\rho_0 .
\eea
For a star in a binary, the main contribution to the potential $\Psi$ comes from the movement of the star along its orbit. It is thus convenient to
decompose $\Psi$ as proposed by Gourgoulhon et al.~\cite{Gourgoulhon2001}:
\bea
\label{eq:defW}
\Psi&=&\Psi_0+W^i x^j \delta_{ij},\\
W^i&=&\left(\frac{\beta^i\phi^4h\gamma_n}{\alpha}\right)_{{\rm CenterNS}}.
\eea
$W^i$ is the inertial velocity at the center of the star, and (\ref{eq:defW}) effectively separates the motion of the star relative to its center from
its orbital motion.

Note that Eq. (\ref{eq:HydroPot}) is derived assuming the existence of an exact helicoidal Killing vector (for more details on the
derivation of (\ref{eq:HydroPot}) from (\ref{eq:continuity}), read Teukolsky ~\cite{Teukolsky1998} and Shibata
~\cite{Shibata1998}). This is, in general, not compatible with our choice of free variables in the XCTS equations. The error we introduce is
most easily seen if we consider the evolution equation for the conformal factor,
\beq
\label{eq:dtphi}
\partial_t \ln{\phi}=\frac{1}{6} \left(-\alpha K+\nabla_k \beta^k\right).
\eeq
For Eq. (\ref{eq:HydroPot}) to be exact, we need $\partial_t \ln{\phi}=0$, while in the XCTS equations we assume that
we are free to choose $K=0$. As nothing guarantees that $\nabla_k \beta^k=0$ --- and in fact, we can check in practice that this term does not vanish --- there is
a contradiction within our equations\footnote{The most natural way to get rid of that contradiction would be to use equation (\ref{eq:dtphi})
as the definition of $K$. The quantity  $\partial_t \ln{\phi}$ would then be a free variable, and could be set to $0$.
However, Pfeiffer showed ~\cite{Pfeiffer2003} that such
a choice makes the operator of the XCTS system noninvertible. Alternatively, inserting (\ref{eq:dtphi}) in an iterative scheme driving
$\partial_t \ln{\phi}$ to 0 seems to be unstable both for BBH ~\cite{Cook2004} and BH-NS binaries.}.

Such approximations are inevitable, as there is no exact equilibrium solution to the binary problem.
In practice, we will see that our numerical scheme is sufficiently accurate that they represent our main source of error. Better choices for $K$, or for our other
free variables, might reduce these errors. However, within the quasiequilibrium formalism, we cannot hope to make them completely disappear. 
In fact, even though the 
contradiction here was shown using the hydrostatic conditions, a quasiequilibrium formulation creates very similar problems in vacuum. 
(A discussion of deviations from quasiequilibrium in BBH binaries can be found in ~\cite{Cook2004}, and the 
amplitude of the time derivative of the conformal factor observed there for irrotational binaries is similar to our results for 
BH-NS binaries.)

Finally, to close our system of equations we need to choose an equation of state (EOS). Here, we will consider a polytropic fluid, 
with polytropic index $\Gamma=2$. The pressure $P$, energy and baryon density $\rho$ and $\rho_0$, internal energy $\epsilon \rho_0$, and
enthalpy $h$ then obey the following relations:
\bea
\label{eq:EOS}
P=\kappa \rho_0^{\Gamma},\\
h=1+\epsilon+\frac{P}{\rho_0}, \\
\rho=(1+\epsilon)\rho_0, \\
\epsilon \rho_0=\frac{P}{\Gamma-1}.
\eea
The method used, however, is independent of the EOS chosen --- as long as, given $h$, we can retrieve $P$, $\rho$, and $\rho_0$.
Indeed, we only use the EOS to reconstruct the matter quantities $\tilde E$, $\tilde J$, $\tilde S$, and $\rho_0$
needed in Eqs. (\ref{eq:XCTSShift}), (\ref{eq:XCTSPhi}), (\ref{eq:XCTSNphi}) and (\ref{eq:HydroPot}) from the enthalpy $h$.
We use a $\Gamma=2$ polytrope as a reasonable first approximation to the nuclear equation of
state, which will allow direct comparison with previous numerical results in Sec.\ \ref{sec:Comparison}.

\subsection{Boundary Conditions}

Building initial data for BH-NS binaries requires us to solve a set of elliptic equations: the
constraints (\ref{eq:XCTSShift}),(\ref{eq:XCTSPhi}), and (\ref{eq:XCTSNphi}) and, in the
case of irrotational binaries, an additional equation for the potential $\Psi$, (\ref{eq:HydroPot}). 
We thus have to provide boundary conditions at infinity and on the BH horizon for the
XCTS variables $\phi$, $\alpha \phi$, and $\beta^i$, and on the surface of the NS for the potential $\Psi$.

At infinity (or, in practice, at $R=10^{10}M$, the outer boundary of our computational domain), we require a flat
Minkowski metric in the inertial frame:
\bea
{\bm\beta}_0=0,\\
\alpha \phi=1 ,\\
\phi=1.
\eea
We excise the BH interior. 
Assuming that the BH is in equilibrium and that the excision surface is
an apparent horizon leads to the set of conditions derived by Cook and Pfeiffer ~\cite{Cook2004}:
\bea
\tilde s^k \tilde \nabla_k \ln{\phi} &=& -\frac{1}{4}\left(\tilde h^{ij}\tilde \nabla_i \tilde s_j - \phi^2 J\right),\\
\beta_{\perp} &=&\beta^i s_i= \alpha,\\
\label{eq:BCShift}
\beta_{\parallel}^i &=& \beta^i-\beta_{\perp}s^i =\Omega_j^{\rm BH} x_k^c \epsilon^{ijk}, 
\eea
where $s^i=\phi^{-2} \tilde s^i$ is the outward unit normal to the surface, $h^{ij}$ its 2-metric, $x_i^c=x_i-c_i$ 
are the Cartesian coordinates relative to its center, $J$ is a projection of the
extrinsic curvature on the excision surface defined in Eq. (28) of ~\cite{Cook2004},
and $\Omega^{\rm BH}$ is a free parameter determining the spin of the black hole. For a corotational BH, $\Omega^{\rm BH}=0$, 
while the value required to obtain a nonspinning black hole is {\it a priori} unknown. 
A good first approximation, suggested in ~\cite{Cook2004}, is $\Omega^{\rm BH}=\Omega$, the orbital angular velocity. 
This choice typically leaves the BH with a spin an order of magnitude lower than in a corotational binary. For better
results, we follow the method introduced by Caudill et al. ~\cite{Caudill2006} for BBH: we iterate over the value of 
$\Omega^{\rm BH}$ to drive the BH spin to zero. This iterative method can be used to generate
a BH of arbitrary spin.

The last boundary condition required on the apparent horizon is only a gauge choice. However,
that choice impacts the amplitude of the deviations from quasiequilibrium ~\cite{Cook2004}.
For conformally flat initial data, we will impose
\beq
\label{eq:BCgauge}
\partial_s \left(\alpha \phi \right)=0,
\eeq
a choice that already gave good results for BBH binaries. We will discuss in Sec.\ \ref{sec:SpinBH} how this condition
is modified when we choose a different conformal metric.

Finally, on the surface of the star, the boundary condition for $\Psi$ can be directly inferred from (\ref{eq:HydroPot}): 
when the density tends towards $0$, we are
left with the equation
\beq
\label{eq:BCPsi}
\tilde\gamma_{ij}\partial_i\Psi\partial_j\rho_0 = \frac{h\gamma_n\beta^i\phi^4}{\alpha}\partial_i\rho_0.
\eeq
As $\tilde \nabla \rho_0$ should be along the normal to the surface of the star,(\ref{eq:BCPsi}) is a boundary condition 
on the normal derivative of $\Psi$.

\subsection{Orbital Angular Velocity}
In the construction of BH-NS initial data, the orbital angular velocity $\Omega$ is, in general, a free parameter. Indeed, 
together with the initial radial velocity, it determines the eccentricity and orbital phase of the orbit. 
Here, we consider binaries
a few orbits before merger, where the trajectory is expected to be quasicircular. As a first approximation, 
we can require force balance at the center of the NS, as proposed by Taniguchi et al.~\cite{Taniguchi2005}:
\beq
\label{eq:forcebalance}
\nabla \ln h = 0.
\eeq
Force balance guarantees that the binary is initially in a circular orbit. As it neglects the infall velocity, it leads
to a slightly eccentric orbit, but still constitutes a good first guess.
Using Eq. (\ref{eq:EulerConstant}), (\ref{eq:forcebalance}) can be written as a condition on the lapse $\alpha$
and the Lorentz factors $\gamma$ and $\gamma_0$:
\beq
\nabla \ln h = \nabla \left(\ln{\frac{\gamma_0}{\alpha \gamma}}\right) =0,
\eeq
or, using the definitions (\ref{eq:Lorentz}) and (\ref{eq:comvel}),
\beq
\label{eq:OmegaOrbit}
\nabla\ln{(\alpha^2-\gamma_{ij}\beta^i\beta^j)}=-2\nabla \ln{\gamma}.
\eeq
Effectively, this is a condition on the orbital angular velocity $\Omega$, if we remember that the shift is
decomposed according to (\ref{eq:shiftdec}). Defining {\bf b} to be the unit-vector along the axis passing through the centers of both compact 
objects, we determine the angular velocity from
\beq
\label{eq:OmegaOrbitAxis}
b^i\nabla_i \ln(\alpha^2 -\gamma_{ij}\beta^i\beta^j) = -2 b^i\nabla_i\ln\gamma. 
\eeq
In theory, the angular velocity appears on both sides of the
equation, but we only write explicitly the left-hand side, keeping $\gamma$ constant. We then check that $\Omega$ 
converges when (\ref{eq:OmegaOrbitAxis}) is inserted in our iterative solving procedure, described in Sec.\ \ref{sec:solver}.

As we only solve (\ref{eq:OmegaOrbit}) along the direction {\bf b}, we
still have to impose force balance along the transverse directions. To do so, we include a correction
term when computing the enthalpy: if $h_0$ is the enthalpy computed from Eq. (\ref{eq:EulerConstant}), we use
as the effective value of $h$
\beq
\label{eq:Drag}
h=h_0\big[1-\left({\bf \nabla}_{\perp}\ln h_0\right)\cdot \left({\bf r}-{\bf c}_{\rm NS}\right)\big],
\eeq
where $\nabla_{\perp}=\nabla- {\bf b}({\bf b}\cdot \nabla)$ and ${\bf c}_{\rm NS}$ is the location of the center of the NS.

This choice drives the maximum of the enthalpy towards ${\bf c}_{\rm NS}$. 
If the equilibrium was exact, ${\bf \nabla}_{\perp}\ln h_0$
would vanish. For our quasiequilibrium binaries, its norm is less than $10^{-6}$.

An alternative method of imposing quasiequilibrium is to use the Komar mass $M_K$. If we have a timelike Killing vector,
then $M_K$ and $\Eadm$, the ADM energy, should be equal. This condition is less convenient to impose during the solution, 
as global quantities like $M_K$ and $\Eadm$ cannot be reliably computed when the constraints are violated. However, we can 
use this equality as a test of our initial data, and verify that $(M_K-\Eadm)$ gets small as we converge.

When we start applying the procedure described by Pfeiffer et al.~\cite{Pfeiffer2008} to reduce the eccentricity of the
system, the situation is slightly different. We then prescribe the value of the orbital angular velocity as well as the initial radial
velocity. Eq. (\ref{eq:OmegaOrbitAxis}) is no longer useful. Instead, we adapt Eq. (\ref{eq:Drag}) 
so that it fixes the position of the star in all three
spatial directions, replacing ${\bf \nabla}_{\perp}$ by ${\bf \nabla}$.

Note that if $\partial_t$ is not an exact Killing vector, the equality between Komar and ADM mass is lost.
We can then use $(M_K-\Eadm)$ only as an indicator of deviations from an exact equilibrium state. For low-eccentricity binaries 
with a nonzero infall velocity, those deviations are significantly larger than when the angular velocity is fixed by Eq. (\ref{eq:OmegaOrbitAxis}),
and the infall velocity set to zero.

\subsection{Observing physical quantities}

We have just seen that, for quasiequilibrium configurations, computing the Komar mass and the ADM energy could be useful 
in finding the optimal angular velocity, 
or to ascertain how far from equilibrium our initial data are. To ensure that our initial configuration has the desired
physical properties, a few additional quantities have to be computed.

First, we want to be able to fix the mass of the compact objects. For a spinning BH, we define the irreducible mass $M_{\rm BH}^{\rm irr}$,
ADM energy in isolation $M_{\rm BH}^{\rm ADM}$, and spin parameter $a_{\rm BH}$,
\bea
M_{\rm BH}^{\rm irr}&=&\sqrt{\frac{A_{\rm AH}}{16\pi}}\\
M_{\rm BH}^{\rm ADM}&=& \frac{(M_{\rm BH}^{\rm irr})^2}{\sqrt{(M_{\rm BH}^{\rm irr})^2-a_{\rm BH}^2/4}}\\
a_{\rm BH}&=&\frac{J_{\rm BH}}{M_{\rm BH}^{\rm ADM}},
\eea
where $J_{\rm BH}$ is the angular momentum of the BH. For the NS, we compute the baryon mass
\beq
\label{eq:Mb}
M_{\rm NS}^b=\int_{\rm NS}\rho_0\phi^6\sqrt{\frac{\tilde\gamma}{1-\gamma_{ij}U^iU^j}}dV.
\eeq
Here, $\tilde \gamma$ is the determinant of the conformal 3-metric $\tilde \gamma_{ij}$.

To check quasiequilibrium, we would like to know the ADM energy and the Komar mass of the system. Measuring the total 
angular momentum is also useful, mainly for comparisons with post-newtonian (PN) predictions or other numerical initial data. Those
quantities are typically defined as integrals on $S_{\infty}$, the sphere at infinity, which is not convenient for 
computations. Integrating by parts, we can transform these expressions into integrals on any sphere $S$ enclosing all matter and 
singularities and, when needed, a volume integral on $V$, the region of our initial slice lying outside of $S$.
Assuming conformal flatness, $K=0$, and no constraint violations, this gives:
\bea
\Eadm &=&-\frac{1}{2\pi}\oint_{S_{\infty}}\delta^{ij}\partial_i \phi dS_j\\ \nonumber
&=&-\frac{1}{2\pi}\bigg(\oint_{S}\delta^{ij}\partial_i \phi dS_j-\frac{1}{8}\int_V\phi^5K_{ij}K^{ij}dV\bigg),
\eea
\bea
M_K&=&\frac{1}{4\pi}\oint_{S_{\infty}}\delta^{ij}\partial_i \alpha dS_j\\ \nonumber
&=&\frac{1}{4\pi}\bigg[\oint_{S}\delta^{ij}\partial_i \alpha dS_j+\int_V\left(\alpha \phi^{-4}\delta^{ik}\delta^{jl}K_{ij}K_{kl}\right. \\ \nonumber
&&\left. -2\phi^{-1}\delta^{ij}\partial_i\alpha\partial_j\phi\right)dV\bigg],\\
\Jadm ^z&=&\frac{1}{8\pi}\oint_{S_{\infty}}\left(xK^{yl}-yK^{xl}\right)dS_l\\ \nonumber
&=&\frac{1}{8\pi}\oint_S\left(xK_{yi}-yK_{xi}\right)\delta^{il}\phi^2dS_l.
\eea
The decomposition into surface and volume integrals is not unique, but we found these expressions convenient, as the contribution of
the volume terms
decreases at least as $1/r$ away from the center of mass, reducing our sensitivity to small numerical
errors at spatial infinity.

To make sure that the axis of rotation of the binary passes through the origin of our numerical grid, we also require
that the ADM linear momentum vanishes. It is computed in a very similar way:
\bea
P_{\rm ADM}^i&=&\frac{1}{8\pi}\oint_{S_{\infty}}K^{ij}dS_j\\ \nonumber
&=&\frac{1}{8\pi}\oint_S\delta^{ik}\delta^{jl}K_{kl}\phi^2 dS_j,
\eea
and our solver moves the position of the BH center so that $P_{\rm ADM}$ is driven to zero.

Finally, when discussing boundary conditions, we have seen that for irrotational binaries the correct value of the parameter
$\Omega_{\rm BH}$ is unknown. We thus need to find the value that makes the BH spin vanish. 
To compute the spin, we use approximate Killing vectors on the apparent horizon, following a method ~\cite{Owen2007} similar to the work
of Cook and Whiting ~\cite{Cook2007}. 

\subsection{Conversion to Physical Units}
\label{sec:units}

In this paper, and in our numerical code, the system of units is based on the arbitrary choice of a unit mass: the ADM
energy of the BH in isolation. Combined with the convention $G=c=1$, this choice is enough to determine all units of interest for BH-NS binaries.
For applications, it is necessary to express results in astrophysical units. In this section, we give the conversion formulas.

We first define the ADM mass of the neutron star $M^{\rm ADM}_{\rm NS}$ as the ADM mass of an isolated NS of baryonic mass $M_{\rm NS}^b$. The
total ADM mass of the binary at infinite separation is then
\beq
M_0=M^{\rm ADM}_{\rm NS}+M^{\rm ADM}_{\rm BH},
\eeq
and the mass ratio is defined as
\beq
{\cal R}=\frac{M^{\rm ADM}_{\rm BH}}{M^{\rm ADM}_{\rm NS}}.
\eeq
Isolated neutron stars of given polytropic index are completely described by their ADM mass and their compactness
\beq
{\cal C}=\frac{M^{\rm ADM}_{\rm NS}}{R_0},
\eeq
where $R_0$ is the areal radius. Furthermore, stars of equal compactness but different masses
are related by a simple scaling law. This can be seen by defining the length scale
\beq
R_{\rm poly}=\kappa^{\frac{1}{2(\Gamma-1)}}
\eeq
and dimensionless quantity
\beq
q=\frac{P}{\rho_0}.
\eeq
The whole problem is then invariant ~\cite{Baumgarte1998} under the transformation
\bea
t'&=&\frac{t}{R_{\rm poly}}\\
{\bf r}'&=&\frac{{\bf r}}{R_{\rm poly}}\\
q'({\bf r}',t')&=&q({\bf r},t).
\eea
In numerical simulations, we can thus retrieve all possible configurations by keeping only ${\cal C}$ and ${\cal R}$ as free
parameters, and choosing $M^{\rm ADM}_{\rm BH}=1$. Systems with different masses but the same neutron star compactness will obey the previous
scaling, with
\beq
R_{\rm poly}=R_{\rm poly}^* \frac{M_0{\cal R}}{1+{\cal R}},
\eeq
and $R_{\rm poly}^*$ the value of $R_{\rm poly}$ when $M^{\rm ADM}_{\rm BH}=1$.
 
We also define
\beq
\zeta({\cal C}) = \frac{M^{b}_{\rm NS}}{M_{\rm NS}^{\rm ADM}},
\eeq
a quantity which, for a given compactness, can easily be obtained from the solution of the Tolman-Oppenheimer-Volkoff (TOV) equation. Then, if the baryon mass of the star 
is expressed in solar masses,
\beq
M^b_{\rm NS}=m_{\rm NS}M_{\odot},
\eeq
the BH ADM energy will be
\beq
M^{\rm ADM}_{\rm BH}=\frac{{\cal R}}{\zeta}m_{\rm NS}M_{\odot}.
\eeq
It is now straightforward to retrieve the meaning of our units of distance and time. A code distance $d$ corresponds to
the physical distance
\beq
D=d\left(\frac{M^{\rm ADM}_{\rm BH}G}{c^2}\right)=d\left(\frac{{\cal R}m_{\rm NS}}{\zeta}\right)\times1.48{\rm km},
\eeq
while a code time $t$ is equal to
\beq
T=t\left(\frac{M^{\rm ADM}_{\rm BH}G}{c^3}\right)=t\left(\frac{{\cal R}m_{\rm NS}}{\zeta}\right)\times4.94\mu {\rm s}.
\eeq
Note, however, that for $D$ to represent an actual physical distance, $d$ has to be the proper separation
\beq
d=\int ds,
\eeq 
and not the coordinate distance on our numerical grid.

In our tests, we choose ${\cal R}=1$ and $\kappa=51.76$, which gives ${\cal C}=0.149$ and $\zeta=1.075$. The conversion
is thus
\bea
D&=&d\left(\frac{m_{\rm NS}}{1.3}\right)\times1.79{\rm km}\\
T&=&t\left(\frac{m_{\rm NS}}{1.3}\right)\times5.97\mu {\rm s}.
\eea

\section{Numerical Methods}
\label{sec:NumMeth}

Turn now to the numerical methods used to solve the initial data problem, and to the way the
solver enforces simultaneously the various constraints on the system derived in Sec.\ \ref{sec:formalism}.
In this paper, we focus on the case of irrotational binaries with no initial radial velocity, even though the solver has also 
been used for single stars, corotational binaries, and infalling binaries. The chosen configuration 
is the most challenging of the four cases: the method for the other cases can be derived by omitting the irrelevant steps
from what we present here.

The core of the problem is the two sets of elliptic equations, the XCTS system (\ref{eq:XCTSShift}),(\ref{eq:XCTSPhi}),
and (\ref{eq:XCTSNphi}), and the irrotational condition on the potential $\Psi$ (\ref{eq:HydroPot}). 
To solve these equations, we use the multidomain spectral 
elliptic solver ({\sc spells}) developed by the Cornell-Caltech collaboration, as described by Pfeiffer et al.~\cite{Pfeiffer2003b}. 
Improvements to {\sc spells} since the publication of ~\cite{Pfeiffer2003b}, mainly the introduction of cylindrical subdomains, 
have increased its efficiency by about a factor of 3. The performance of the solver on distorted subdomains ---
such as a subdomain with a boundary chosen to follow the surface of the neutron star --- has also been improved, allowing us
to solve the initial data problem in the presence of matter without Gibbs oscillations at the surface of the
star.

{\sc spells} has already been used successfully
to solve the XCTS system for BBH binaries ~\cite{Pfeiffer2003,Pfeiffer2008}. 
Here, when solving
for the XCTS variables, we consider the matter terms as fixed, while in (\ref{eq:HydroPot}), only the potential
$\Psi$ is variable. We will detail in Sec.\ \ref{sec:solver} how to combine the two groups of 
equations, as well as the additional conditions of force balance (\ref{eq:forcebalance}), vanishing ADM
linear momentum and BH spin, and known BH and NS masses. But first we discuss some aspects of the solution of the 
elliptic equations themselves: the numerical grid, and specifics of the irrotational potential equation.

\subsection{Domain Decomposition}

\subsubsection{Numerical Grid}
\label{sec:numgrid}

The flexibility of the multidomain method used by {\sc spells} allows us to use relatively complex subdomain decompositions, adapting
the numerical grid to the geometry of the problem at hand. It also makes it possible to solve directly the whole XCTS system as a 
single set of nonlinear equations, without further decomposition of the XCTS variables, and using a relatively low number of grid points.

\begin{figure}
\includegraphics[scale=0.35]{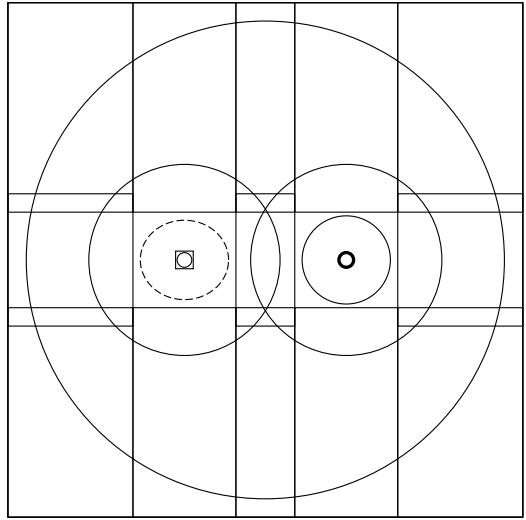}
\caption{\label{fig:Sd} Subdomain decomposition close to the compact objects, in the equatorial plane. The apparent horizon of the BH (right) is an inner boundary of 
the numerical domain, while the surface of the NS (dashed line) is the boundary between
the two spherical shells on the left. }
\end{figure}

For binaries in {\sc spells}, we build the numerical grid from 14 subdomains, as follows (see Fig.\ \ref{fig:Sd}):
\begin{itemize}
\item Around the BH, we use two concentric spherical shells and require their innermost boundary to be an apparent horizon.
\item The neighborhood of the NS is covered by an outer spherical shell with inner boundary mapped to the surface of the neutron star.  This outer spherical shell touches an inner spherical shell which covers the whole neutron star, except a small region at the center.  To avoid having to deal with regularity conditions at the center of a full sphere, the central region is covered by a cube overlapping the inner spherical shell.
\item Three rectangular parallelepipeds cover the region surrounding the axis passing through the centers of the compact
objects: one between the BH and the NS, and one on each side of the binary.
\item Five cylindrical shells around the same axis cover the intermediate field region. Their innermost boundary is, for
three of them, within the parallelepipeds, and for the other two, within the outer shell surrounding each compact object.
\item The far-field region is covered by a spherical shell, with a $1/r$ coordinate mapping allowing us to place
the outer boundary at spatial infinity (or, in practice, at $R=10^{10}M$).
\end{itemize}
At the second highest resolution, which we use as a reference to estimate the accuracy of the solution, the cube
at the center of the star has 11$\times$11$\times$11 collocation points, the spherical
shells around the compact objects have 19$\times$18$\times$36 points, the parallelepipeds 13$\times$20$\times$20 points, the cylinders 14$\times$15$\times$13 
(15 in the angular direction) or 14$\times$15$\times$20 (the higher resolution for the subdomains closer to the compact objects),
and the outer sphere 12$\times$10$\times$20. For comparison, the numerical grid used in ~\cite{Taniguchi2008} is built out of spherical shells
with resolution 41$\times$33$\times$32 or 49$\times$37$\times$36 around the black hole, and 25$\times$17$\times$16 around the neutron star.

To make convergence tests, we will need a single measure of the resolution used. For a domain decomposition
using subdomains with different basis functions and number of collocation points, this definition is certainly not unique.
We will use
\beq
\label{eq:defN}
N^{1/3}=\left(\sum_{\rm Subdomains} N_i\right)^{1/3},
\eeq
where $N_i$ is the number of collocation points in subdomain $i$. For our second highest resolution, $N^{1/3}=44.0$, while
for Ref. ~\cite{Taniguchi2008} $N^{1/3}>78.7$. 

\subsubsection{Surface Fitting}

Discontinuities in variables within a subdomain spoil spectral convergence. The surface of the star is a discontinuity,
so we make it the boundary between two subdomains. (Note, however, that it is possible to reach a good level of precision --- of the order
of the error coming from deviations from quasiequilibrium --- simply by including the surface in the interior of a thin spherical shell.)

The surface of the star is approximated by an expansion in spherical harmonics,
\beq
\label{eq:rexp}
R_{\rm surf}=\sum_{lm} c_{lm} Y^{lm}(\theta,\phi),
\eeq
where the center of the star, as defined in Eq. (\ref{eq:forcebalance}), is the origin of the spherical coordinates. To determine the 
coefficients $c_{lm}$, we solve the equation $h(R_{ij},\theta_i,\phi_j)=1$ along each collocation direction $(\theta_i,\phi_j)$ of the 
numerical grid. Then, we project onto spherical harmonics the function $R(\theta,\phi)$ defined by its values $R_{ij}$ in each collocation direction. 

To avoid Gibbs oscillations, we force the surface to be at the boundary between two spherical shells, $S_0$ and $S_1$. This is done by
a coordinate transformation $R \rightarrow R'$ fixing the radius of the common boundary between $S_0$ and $S_1$ to be the given
function $R_{\rm bound}(\theta,\phi)$. This function is expanded in spherical harmonics, and will
be equal to $R_{\rm surf}$ when the solver converges, as explained in Sec.\ \ref{sec:solver}.
If $S_0$ is defined in the original coordinates by $R_0<R<R^*$, and $S_1$ by $R^*<R<R_1$, the map is, in $S_0$,
\beq
\label{eq:coordmap}
R'(\theta,\phi) = \frac{R_{\rm bound}(\theta,\phi)-R_0}{R^*-R_0}(R-R_0)+R_0, 
\eeq
while in $S_1$ we have
\beq
R'(\theta,\phi) = \frac{R_{\rm bound}(\theta,\phi)-R_1}{R^*-R_1}(R-R_1)+R_1. 
\eeq
The exact value of $R^*$ is not important, as long as $R_0<R^*<R_1$. However, having $R^*\sim R_{\rm bound}$ is usually convenient,
as it leads to $R\sim R'$. 

The validity of the $Y_{lm}$ expansion is evaluated by observing the convergence of the coefficients
$c_{lm}$ as the resolution increases. Results for a test irrotational binary are discussed in Sec.\ \ref{sec:ConvBin}.

\subsection{Irrotational flow}
\label{sec:Irrot}

Once the domain decomposition has been chosen, the XCTS equations can be solved without further modification. 
The irrotational equation (\ref{eq:HydroPot}), however, has specific problems that require further attention.

First, the coefficient of the leading order term --- the Laplacian --- vanishes on the surface of the star. As the equation is 
preconditioned by the inverse of a finite difference approximation of the flat Laplacian, convergence will become extremely
poor close to the surface, where (\ref{eq:HydroPot}) is very different from Laplace's equation.
We thus change the preconditioning operator from an approximation of $-\nabla^2 u$ to an approximation of
$-\rho_0 \nabla^2 u+u$. The leading order term will then be properly represented within the star, while, when the
density decreases, the operator becomes the identity and no preconditioning is done.

Another problem is related to the inconsistencies in the quasiequilibrium formulation, already discussed in Sec.\ \ref{sec:hydroeq}.
Indeed, we know that, for a perfect equilibrium, Eq. (\ref{eq:HydroPot}) will admit an infinite number of solutions (the
potential is only defined up to a constant term). But, if we have instead a quasiequilibrium situation, Eq. (\ref{eq:HydroPot})
is not an exact representation of the continuity equation anymore. And nothing guarantees that a solution even exists. We found 
in practice that when using
Eq. (\ref{eq:HydroPot}) as written, the convergence of the solver stops before we reach an acceptable
precision.

Different solutions to this problem were tried, involving small modifications of Eq. (\ref{eq:HydroPot}). Here small 
means ``at most of the order of the deviations from quasiequilibrium.'' The results presented here were obtained by
replacing $K$ in (\ref{eq:HydroPot}) by the value required to ensure that $\partial_t \ln \phi=0$ using Eq. (\ref{eq:dtphi}).
Of course, this does not solve the inconsistency --- $K$ is still set to $0$ in the XCTS equations --- but it guarantees that
Eq. (\ref{eq:HydroPot}) has a solution, allows the system to converge, and does not introduce any new source of error.

Another method, mathematically less satisfactory but leading to equivalent results, is to allow for a small correction in
(\ref{eq:HydroPot}), for example, by adding the mean value of the potential, $\tilde{\Psi}$, to the boundary condition (\ref{eq:BCPsi}) and
requiring that $\tilde{\Psi}$ is driven to zero (or, in practice, the small value required to counter the error coming from
our choice of $K$) as we converge.

\subsection{Building quasiequilibrium binaries}
\label{sec:solver}

As discussed previously, knowing how to solve each set of elliptic equation is only part of the problem. Here, we outline
how the solver links all of the requirements together and ensures convergence towards a solution representing 
the desired physical situation.

At a fixed resolution, we solve according to the following algorithm:
\begin{enumerate}
\item Solve the XCTS system (\ref{eq:XCTSShift}), (\ref{eq:XCTSPhi}), and (\ref{eq:XCTSNphi}), with fixed conformal matter quantities
$\tilde E$, $\tilde S$ and $\tilde J^i$. The new value of the
XCTS variables is determined by the relaxation formula $u_{n}=(1-\lambda)u_{n-1}+\lambda u^*$, where $\lambda$ is an
arbitrary parameter (we typically use $0.3$) and $u^*$ the value of $u$ found by solving the XCTS equations. In fact,
knowing that we will use a relaxation formula, we do not even solve the equations exactly at each iteration; 
an approximate solution
is good enough, and saves a lot of computer time.
\item Impose symmetry across the equatorial plane (this step is not required, but we know that this symmetry should be
respected, and enforcing it strictly accelerates convergence).
\item Evaluate the position of the surface of the star, $R^n_{\rm surf}$, and compare it to the evaluation made
during the previous iteration, $R^{n-1}_{\rm surf}$. If both agree within a certain precision --- we use the condition 
$||R^n_{\rm surf}-R^{n-1}_{\rm surf}||_2<0.1 ||R^n_{\rm surf}-R_{\rm bound}||_2$, where $R_{\rm bound}$ is the function used in the mapping
(\ref{eq:coordmap}) --- modify the numerical grid by setting $R_{\rm bound}=R^n_{\rm surf}$.
\item Compute the ADM linear momentum ${\bf P}^n_{\rm ADM}$, and compare it to the value computed during the
previous iteration, ${\bf P}^{n-1}_{\rm ADM}$. If $||{\bf P}^n_{\rm ADM}-{\bf P}^{n-1}_{\rm ADM}||<0.1\times||{\bf P}^n_{\rm ADM}||$, move the center of the BH.
The change in the position of the center, $\delta {\bf c}$, is chosen so that,
if the system was Newtonian, the total linear momentum would vanish: $\delta {\bf c} \times {\bf \Omega}={\bf P}^n_{\rm ADM}$.
We also change the radius of the excision surface (the inner boundary of the shells around the BH) to drive $M^{\rm ADM}_{\rm BH}$ 
to its desired value.
\item Solve Eq. (\ref{eq:OmegaOrbitAxis}) to find the new angular velocity.
\item Get the spin of the BH, and change the parameter $\Omega^{\rm BH}$ in the boundary condition (\ref{eq:BCShift}) to drive
the spin to 0 --- or any other desired value, if the BH is not irrotational.
The new value of $\Omega^{\rm BH}$ is chosen by linear interpolation, using the last two values of the spin.
\item Determine the constant in the Euler first integral (\ref{eq:EulerConstant}) so that the baryon mass of the NS (\ref{eq:Mb})
is set to its target value.
\item Apply correction (\ref{eq:Drag}) to the value of the enthalpy.
\item Solve the irrotational equation (\ref{eq:HydroPot}) for $\Psi$. The new value of $\Psi$ is determined using the same relaxation
formula as for the XCTS variables.
\item If the desired precision has not been reached, go back to 1.
\end{enumerate}

From this description, it is clear that the accuracy of the results depends on the convergence of the many parameters updated
during the iterative procedure. We will discuss in Sec.\ \ref{sec:ConvBin} various tests verifying that they all reach
an acceptable precision.

\section{Tests and Results}
\label{sec:NumTests}

As mentioned earlier, the main motivation to build a code generating BH-NS initial data using a multidomain spectral
method is the possibility of rapidly reaching high levels of precision. As an example, we will focus on a sequence of irrotational, 
equal-mass BH-NS binaries. In Sec.\ \ref{sec:ConvBin}, we show through convergence tests that, over a large range of 
initial separations likely to be chosen as starting points for future evolutions, we can construct initial data with enough 
precision to resolve deviations from quasiequilibrium. Trying to reach higher precision, even if 
mathematically possible, would be of little interest: the additional information would not be physically meaningful.

We then turn, in Sec.\ \ref{sec:Comparison}, to another interesting test of our results: comparing them to a similar
sequence generated by Taniguchi et al.~\cite{Taniguchi2008}, and to predictions from the 3PN approximations computed by Blanchet
~\cite{Blanchet2002}, as well as Mora and Will ~\cite{Mora2005}. With accurate estimates of our errors, we discuss how far deviations 
of the numerical results
from the 3PN approximations can be trusted, and their potential interpretation.

Finally, we end this section with a discussion of two different types of initial configurations: binaries built using a modified Kerr-Schild
conformal metric to construct systems with a spinning BH, and binaries with an initial radial velocity, which can be used to
generate systems with low-eccentricity orbits.

\subsection{TOV Star}
\label{sec:TOV}

\begin{table}
\caption{Domain decomposition for a single TOV star. For spherical shells, the three numbers
denote the resolution in radial, polar and azimuthal directions.}
\label{tab:resTOV}
\begin{tabular}{|c|c|c|c|}
\hline
 & Central Cube & Inner Shells & Outer Shell \\
 \hline
 R0 & 7$\times$7$\times$7 & 7$\times$6$\times$12 & 8$\times$6$\times$12 \\
 \hline
 R1 & 9$\times$9$\times$9 & 10$\times$9$\times$18 & 9$\times$7$\times$14 \\
 \hline
 R2 & 11$\times$11$\times$11 & 13$\times$12$\times$24 & 10$\times$8$\times$16 \\
 \hline
 R3 & 13$\times$13$\times$13 & 16$\times$15$\times$30 & 11$\times$9$\times$18 \\
 \hline
\end{tabular}
\end{table}

\begin{figure}
\includegraphics[scale=0.48]{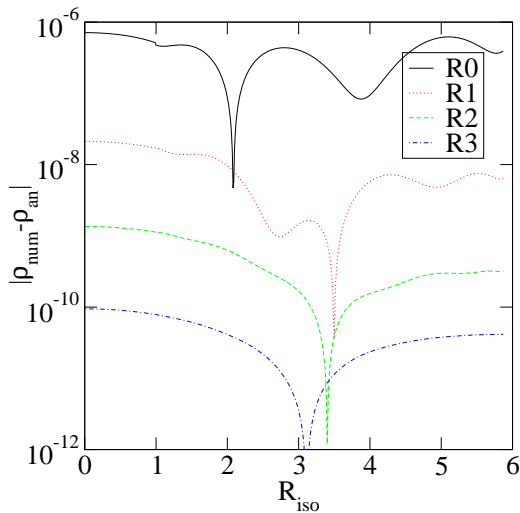}
\caption{\label{fig:TOVStar} Error in the energy density for an isolated NS
as a function of the isotropic radius $R_{\rm iso}$.
The reference configuration is obtained by numerical integration of the TOV equations.
The spikes in the error function are due to a change in the sign of $(\rho_{\rm num}-\rho_{\rm an})$}
\end{figure}

Before tackling binaries, we test our algorithm on an isolated, nonrotating NS. This effectively means that only steps 1, 2, 3, and 7
of our solution procedure are not trivial. Although the position of the surface is known analytically, for the
purpose of this test we rely on the iterative surface fitting method to find it.
An ``exact'' solution is easily computed by direct integration of the TOV equations. We compared the central density, ADM mass, Komar mass, and central lapse: 
all converge exponentially with resolution.

Figure \ref{fig:TOVStar} shows the difference between the exact and computed density profiles. We can see
that the spectral convergence of the error holds at all radii.

For this simple case, the domain decomposition consists of just a cube covering 
the center
of the star, two spherical shells whose common boundary matches the surface, and a third shell with an
$1/r$ mapping extending to $R=10^{10}M$. The resolutions R0 to R3 used in the test are described in 
Table \ref{tab:resTOV}.

\subsection{Irrotational binaries}
\label{sec:ConvBin}

To test the performance of our solver for binary systems, we use the iterative method from Sec.\ \ref{sec:solver} to construct
a sequence of equal-mass, irrotational binaries. The NS has an ADM mass in isolation of 1 (in code units: see Sec.\ \ref{sec:units} 
for a conversion in astrophysical units), and a parameter $\kappa=51.76$, leading to a compaction similar to that used in 
Ref. ~\cite{Taniguchi2008}, Table IV. Our results are detailed in Table \ref{tab:MR1}. 

\begin{table}
\caption{Sequence of irrotational, equal-mass BH-NS binaries. We give here the coordinate distance between the centers of the two
compact objects $d$, the orbital angular velocity $\Omega$, the binding energy $E_b$,
the angular momentum $J$, and the difference between Komar and ADM energies.\\
For the four closest configurations, marked by an asterisk, the numerical error estimated from the convergence of energy measurements
is larger than the deviations from quasiequilibrium, approximated by $\delta M=M_K-E_{\rm ADM}$, so that $\delta M$ might not be
resolved. The error in $E_b$ reaches about $5\times10^{-5}$ at $d/M_0=8.406$, an order of magnitude larger than at
$d/M_0=9.007$. }
\label{tab:MR1}
$\begin{array}{|c|c|c|c|c|}
\hline
\frac{d}{M_0} & \Omega M_0 & \frac{E_b}{M_0} & \frac{J}{M_0^2} & \frac{|E_{\rm ADM}-M_K|}{M_0} \\
\hline
18.505 & 0.01169 & -6.1490 \times 10^{-3} & 1.19460 & 7.7\times 10^{-7} \\
16.506 & 0.01377 & -6.8103 \times 10^{-3} & 1.14241 & 9.5\times 10^{-7} \\
14.506 & 0.01653 & -7.6289 \times 10^{-3} & 1.08815 & 1.4\times 10^{-6} \\
12.506 & 0.02037 & -8.6634 \times 10^{-3} & 1.03177 & 2.0\times 10^{-6} \\
11.506 & 0.02288 & -9.2879 \times 10^{-3} & 1.00284 & 2.6\times 10^{-6} \\
10.507 & 0.02596 & -1.0002 \times 10^{-2} & 0.97353 & 3.4\times 10^{-6} \\
9.507  & 0.02981 & -1.0821 \times 10^{-2} & 0.94408 & 4.4\times 10^{-6} \\
9.257  & 0.03092 & -1.1043 \times 10^{-2} & 0.93675 & 5.0\times 10^{-6} \\
9.007  & 0.03211 & -1.1273 \times 10^{-2} & 0.92947 & 5.4\times 10^{-6} \\
8.857  & 0.03285 & -1.1416 \times 10^{-2} & 0.92514 & 5.6\times 10^{-6}*\\
8.757  & 0.03337 & -1.1509 \times 10^{-2} & 0.92225 & 6.0\times 10^{-6}*\\ 
8.557  & 0.03445 & -1.1706 \times 10^{-2} & 0.91656 & 6.6\times 10^{-6}*\\
8.406  & 0.03530 & -1.1853 \times 10^{-2} & 0.91237 & 8.2\times 10^{-6}*\\
\hline
\end{array}$
\end{table}

We look at three different sources of error.

\begin{itemize}
\item The iterative procedure. To estimate that error, we study the convergence, at fixed resolution, 
of all the parameters changing between iterations.

\item Truncation errors. We observe the convergence of the solution with the number of collocation points by solving each configuration
at four different resolutions, R0 to R3, as detailed in Table \ref{tab:resbin}.The second highest resolution, R2, 
is our standard numerical grid, as defined in Sec.\ \ref{sec:numgrid}, and the highest resolution, R3, is used as an 
approximation of the exact solution.

\item Deviations from equilibrium. We know that the quasiequilibrium formalism contains intrinsic contradictions. 
A useful estimate of the error
thus created is the difference between the Komar and ADM energies. In the presence of an exact timelike Killing vector, both
would be equal, but here the difference can be seen as an indication of how far from equilibrium we are.
\end{itemize}
All the graphs presented in this section correspond to a binary with rescaled coordinate separation $d/M_0=11.507$. 
A summary of our results for the whole sequence is in Table \ref{tab:MR1}. Typically, the numerical error
rises as the separation decreases. The difference between Komar and ADM mass can be resolved up to
$d/M_0=9$. Numerical errors then start to increase rapidly to reach, for our closest binary, values around 
$5\times10^{-5}$. By
that point, the solver does not converge at resolution R3 anymore, and we thus use R1 as our reference and R2 as an
estimate of the exact solution.

\begin{table}
\caption{Domain decomposition for binary systems. A description of the different subdomains
can be found in Sec.\ \ref{sec:numgrid}. The three numbers denote the resolution in radial, 
polar,and azimuthal directions for spherical shells, and in radial, polar, and axial directions 
for the cylinders. The cylinders have two different resolutions (HR/LR), the highest being
used for the two subdomains directly surrounding one of the compact object. Finally, for the
parallelepipeds, the first number corresponds to the resolution along the axis passing through the
centers of both compact objects.}
\label{tab:resbin}
\begin{tabular}{|c|c|c|c|c|c|}
\hline
 & Cube & Inn. Shells & Out. Shell & Parall. & Cyl.(HR/LR).\\
 \hline
 R0 & 9$\times$9$\times$9 & 13$\times$12$\times$24 & 8$\times$6$\times$12 & 9$\times$12$\times$12 & 10$\times$9$\times$12/9 \\
 \hline
 R1 & 10$\times$10$\times$10 & 16$\times$15$\times$30 & 10$\times$8$\times$16 & 11$\times$16$\times$16 & 12$\times$12$\times$16/11 \\
 \hline
 R2 & 11$\times$11$\times$11 & 19$\times$18$\times$36 & 12$\times$10$\times$20 & 13$\times$20$\times$20 & 14$\times$15$\times$20/13 \\
 \hline
 R3 & 12$\times$12$\times$12 & 22$\times$21$\times$42 & 14$\times$12$\times$24 & 15$\times$24$\times$24 & 16$\times$18$\times$24/15\\
 \hline
\end{tabular}
\end{table}

\subsubsection{Convergence of the iterative procedure}

\begin{figure}
\includegraphics[scale=0.48]{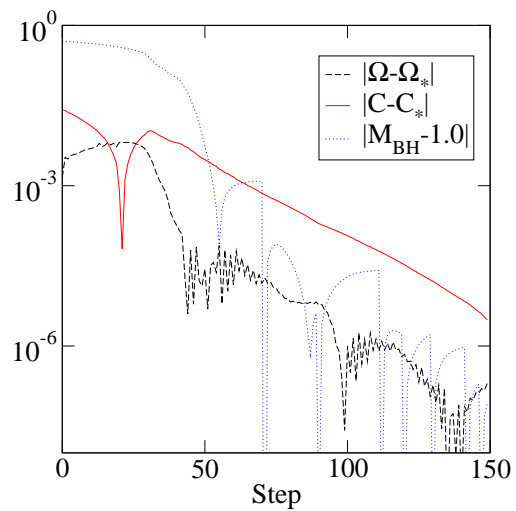}
\caption{\label{fig:Lvl0Conv} Convergence of the angular velocity, the
Euler constant (which controls the mass of the star) and the mass of the BH
while iterating at the lowest resolution R0 for an equal-mass binary with
initial separation $d/M_0=11.507$. The values plotted are the differences from the
final results at the highest resolution R3. One step is defined as a passage from point 1
to point 10 in the iterative procedure described in Sec.\ \ref{sec:solver}}
\end{figure}

\begin{figure}
\includegraphics[scale=0.48]{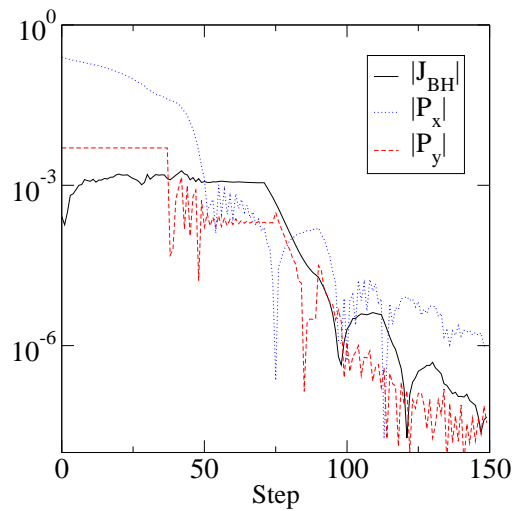}
\caption{\label{fig:Lvl1SpinP} Convergence of the spin of the BH $J_{BH}$ 
and the total linear momentum $P_{ADM}$ at our
lowest resolution R0 for an equal-mass binary with initial separation
$d/M_0=11.507$. }
\end{figure}

To verify the convergence at fixed resolution, observe Figs.\ \ref{fig:Lvl0Conv} and \ref{fig:Lvl1SpinP}. 
The iterative procedure converges if all parameters modified within one step converge, while
the residuals from the two elliptic solves (i.e., the constraint violations and the deviations of the fluid from an
irrotational configuration) vanish. In Fig.\ \ref{fig:Lvl0Conv}, we show the evolution of three of these parameters while
iterating at our lowest resolution R0: the angular 
velocity $\Omega$, derived from the Eq. (\ref{eq:OmegaOrbitAxis}), the constant in the Euler first 
integral (\ref{eq:EulerConstant}), which controls the mass
of the NS, and the areal mass of the BH, controlled by the radius of the excision surface. The difference between
the parameter at a given step and its final value at the highest resolution is shown. We see that, even though
the resolution is low, all parameters converge to a relative precision below $10^{-5}$. At the reference resolution R2,
the relative precision is better than $10^{-7}$.

In addition to the overall convergence, Fig.\ \ref{fig:Lvl0Conv} shows abrupt changes, especially in the evolution
of the BH mass. These can easily be understood if we remember how the mass of the BH is fixed: the radius of the excision
boundary is modified whenever the linear ADM momentum converges. We then change our numerical grid and the location of
the apparent horizon. Every time we regrid,
the BH mass will at first be very close to its desired value, then reach a new equilibrium when the system adapts 
to its new boundary condition. The mass just before regridding --- when the error is maximal --- is thus the best estimate 
of our precision.

We also monitor the evolution of a number of quantities that should tend 
towards zero as the system
converges: the total linear momentum (to ensure that the axis of rotation passes through the origin of our coordinate 
system), the BH spin (as we want irrotational binaries), the quantity $\nabla_{\perp} \ln h$ in Eq.
(\ref{eq:Drag}), and the L2 norm of modes violating the equatorial symmetry (before we manually impose it). 
The last converges
quickly to relative precisions of order $10^{-7}$, and down to about $10^{-10}$ at resolution R2, while the behavior 
of $P_{\rm ADM}$ and $J_{\rm BH}$ is shown on 
Figure \ref{fig:Lvl1SpinP}. As in Fig.\ \ref{fig:Lvl0Conv}, we plot the evolution at our lowest resolution, R0. 
We observe rapid convergence, with once more some oscillations due to the occasional modification of the numerical grid.
At the reference resolution R2, both $P_{\rm ADM}$ and $J_{\rm BH}$ vanish to a precision better than $10^{-9}$.
From Figs.\ \ref{fig:Lvl0Conv} and \ref{fig:Lvl1SpinP}, we can thus safely consider that the iterative method detailed
in Sec.\ \ref{sec:solver} does indeed converge at fixed resolution.

The last parameter, $\nabla_{\perp} \ln h$ (not plotted), does not however completely vanish, even at our highest resolution. In fact,
it converges rapidly towards a fixed, small value of order $10^{-7}$. This is most likely because the equilibrium
is not perfect --- and, indeed, when the deviations from exact equilibrium increase, so does the final value of 
$\nabla_{\perp} \ln h$.


\subsubsection{Spectral convergence of the solution}

Having established that the iterative procedure works as intended, we turn to an estimate of the precision of the initial data
obtained, that is, the differences between the solutions at different resolutions. As we
use a spectral representation, we expect exponential convergence of all variables. We report the convergence of the constraint 
violations, the performance of the surface fitting method, and the convergence of a set 
of measured global quantities ($\Eadm$, $J_{\rm ADM}$, $M_K$, and the position of the BH center ${\bf c}_{\rm BH}$).

\begin{figure}
\includegraphics[scale=0.48]{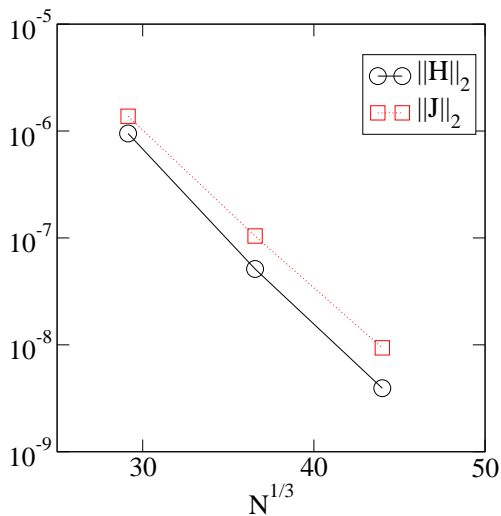}
\caption{\label{fig:Constraints} Convergence of the Hamiltonian and
momentum constraints
with resolution for an equal-mass binary at a separation $d/M_0=11.507$.}
\end{figure}

Fig.\ \ref{fig:Constraints} shows the residual of the elliptic equations corresponding to the Hamiltonian 
and momentum constraints. At the end of an elliptic solve at any given resolution, it should vanish
at all collocation points. In order to obtain a meaningful estimate of the error, we thus evaluate the residual
on the numerical grid corresponding to the next higher resolution. The exponential
convergence is clearly seen, and we can deduce from Fig.\ \ref{fig:Constraints} that the
norm of the constraints at resolution R2 is around $10^{-8}$.

\begin{figure}
\includegraphics[scale=0.48]{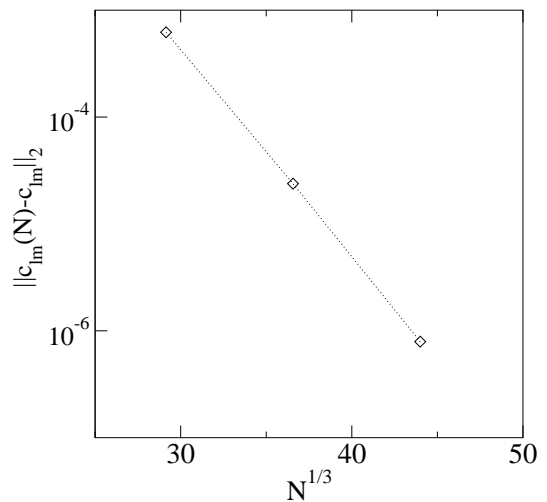}
\caption{\label{fig:SurfaceError} 
Convergence of the surface fitting method measured as the evolution of the error
in the coefficients of the expansion of $R_{\rm surf}$ in spherical harmonics, computed here
as the difference with our results at our highest resolution. }
\end{figure}

\begin{figure}
\includegraphics[scale=0.48]{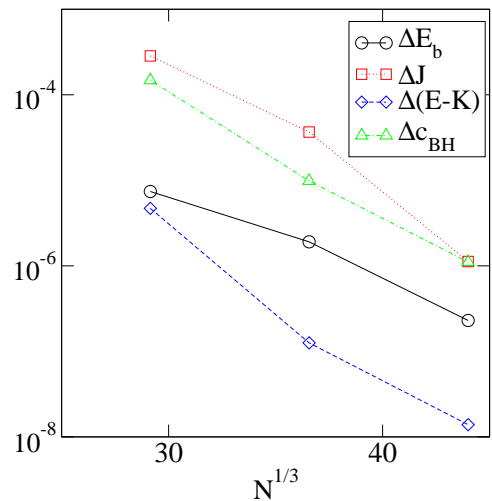}
\caption{\label{fig:ADMquantities} Convergence of the ADM energy, ADM angular
momentum, difference between Komar and ADM energies, and position of the 
center of the BH with resolution for an equal-mass binary at a separation 
$d/M_0=11.507$. 
We plot $\Delta E_b=(E_b-E_b^*)/M_0$, $\Delta J=(J-J^*)/M_0^2$, $\Delta (E-K)=(\delta M-\delta M^*)/M_0$, 
and $\Delta c=||{\bf c}_{\rm BH}-{\bf c}_{\rm BH}^*)||_2$, where the reference results
$E_b^*$, $J^*$, $\delta M^*$ and ${\bf c}_{\rm BH}^*$ are those at resolution R3 ($N^{1/3}=51.5$).\\
The difference $\delta M$ between $M_K$ and $E_{\rm ADM}$, an indication
of how close to equilibrium the system is, reaches $2.6\times10^{-6}$ at the highest
resolution. This is significantly larger than the estimated error in either $E_{\rm ADM}$,
or $\delta M$ shown in the figure.}
\end{figure}

The performance of the surface fitting method can be evaluated from Fig.\ \ref{fig:SurfaceError}, where we 
show the convergence of the surface at different resolutions. The error is estimated by the L2 norm of the 
difference between the coefficients of the expansion in spherical harmonics (\ref{eq:rexp}) at the current resolution 
and their final values at our highest resolution. The exponential convergence allows us to
easily estimate the error in the position of the surface. For this configuration the position of the surface is known within
better than $10^{-6}$ code units. 
For highly distorted stars however, this error becomes significant, and provides the easiest way to check 
during the computation whether the angular resolution is high enough or not.

Finally, in Fig.\ \ref{fig:ADMquantities} we show the 
convergence of the measured ADM energy and angular momentum with resolution. For both quantities, the reference for 
comparison is the value measured at the highest resolution R3. We see good convergence over 2 orders of magnitude.
Similar figures can be obtained for different binary separations --- though as discussed earlier, our ability to solve
accurately at high resolution decreases when the star becomes too distorted.

\subsubsection{Deviations from equilibrium}

Also in Fig.\ \ref{fig:ADMquantities}, we plot the convergence of the position of the BH center, which confirms that the center of
the numerical grid is indeed the center of rotation of the system, and the convergence of the difference between the ADM
and Komar energies $\delta M$, a measure of the deviation from quasiequilibrium. We see that this difference is resolved to a very
high precision, much lower than its actual value of $2.6\times10^{-6}$. As the ADM energy itself is also resolved to a
precision significantly better than $10^{-6}$, we see that our main sources of imprecision are the 
inconsistencies inherent in the quasiequilibrium approximation.

Both the numerical errors and the deviations from quasiequilibrium increase as the separation decreases, but, as
the star approaches its mass-shedding limit, the numerical error increases much more rapidly. As previously mentioned, they are 
roughly comparable
for a rescaled coordinate separation $d/M_0=9$. The decrease in performance at lower separations is not, however,
a serious problem. By that point, the radial velocity of any real binary will already be significant, and so would other
deviations from the idealized quasiequilibrium state. Any evolution looking for such levels of precision should 
probably start from a larger separation. Results at small coordinate separations are, however, interesting for more
qualitative predictions. For example, Taniguchi et al. ~\cite{Taniguchi2008} use them to determine which configurations
are likely to reach the innermost stable orbit before the star gets disrupted. We will thus keep them as useful
approximations, without expecting the same precision as for more widely separated objects.

\subsection{Comparison with previous results}
\label{sec:Comparison}
As a last test of our code, we compare the initial data generated using the iterative method described in
Sec. \ref{sec:solver} to 3PN approximations and previous numerical results. For these comparisons, we use
the sequence of equal-mass, irrotational binaries detailed in Table \ref{tab:MR1}.
The 3PN values were obtained in the point-mass, circular orbit approximation by Blanchet ~\cite{Blanchet2002}. We also use
results from Mora and Will ~\cite{Mora2005} to take into account eccentricity and finite size effects.
For the numerical comparison, we use the data from Table IV of Taniguchi et al. ~\cite{Taniguchi2008}.
These last results are given to 3 significant digits, the actual precision being unknown to us. 
Their error in the quasiequilibrium
condition --- our sole basis for comparison --- is, at most separations, around an order of magnitude higher than what we observe 
in our initial data.
This error is, however, small enough to allow comparisons of both numerical results with the 3PN approximations. 

Four different models are compared. The first corresponds to the results of Blanchet~\cite{Blanchet2002}, where the orbits
are circular and the compact objects are modeled as point masses. The second adds finite size effects to the model. 
Most corrections made by Mora and Will~\cite{Mora2005} to the point-mass model vanish in the case of an irrotational binary, and only the tidal
effects add a significant contribution. We compute them according to Eq. (3.6a) of their work. The last two models
include some eccentricity. The exact eccentricity of our initial data is, in general, unknown. However, we can get reasonable
estimates from evolutions starting at separation $d/M_0=12$. We will give the 3PN results for
binaries with an eccentricity $e=0.01$. At a given eccentricity, the binding energy and ADM momentum reach extrema at the pericenter
and the apocenter. We thus present the 3PN results at those two points, giving an order of magnitude estimate
of the influence of the eccentricity. A summary of the parameters chosen for the four models is given in Table \ref{tab:3PN}.\\

\begin{table}
\caption{Choice of 3PN models used as references. The eccentricity {\it e} is defined as in ~\cite{Mora2005}, Eq. (2.3)}
\label{tab:3PN}
\begin{tabular}{|c|c|c|c|c|}
\hline 
& Source & Finite size & {\it e} & Orbital pos.\\
\hline
3PN-B & Blanchet~\cite{Blanchet2002} & No & 0 & --- \\
3PN-M0 & Mora and Will~\cite{Mora2005} & Yes & 0 & --- \\
3PN-MP &  Mora and Will~\cite{Mora2005} & Yes & 0.01 & Pericenter \\
3PN-MA &  Mora and Will~\cite{Mora2005} & Yes & 0.01 & Apocenter \\
\hline
\end{tabular}
\end{table}

\begin{figure}
\includegraphics[scale=0.48]{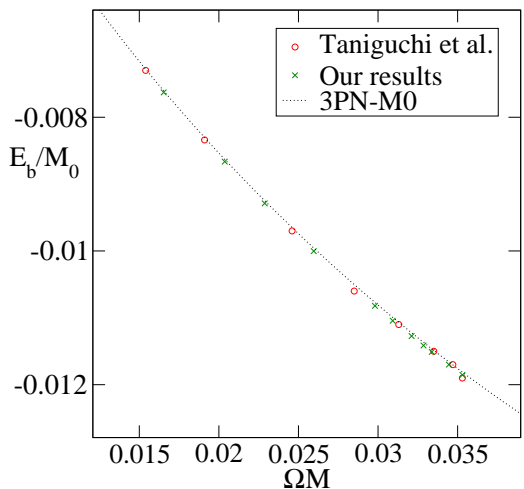}
\caption{\label{fig:Eb} Binding energies of equal-mass binaries for initial data
from our solver, from Taniguchi et al. ~\cite{Taniguchi2008}, and from 3PN
predictions for model 3PN-M0 (see Table \ref{tab:3PN}). }
\end{figure}

\begin{figure}
\includegraphics[scale=0.48]{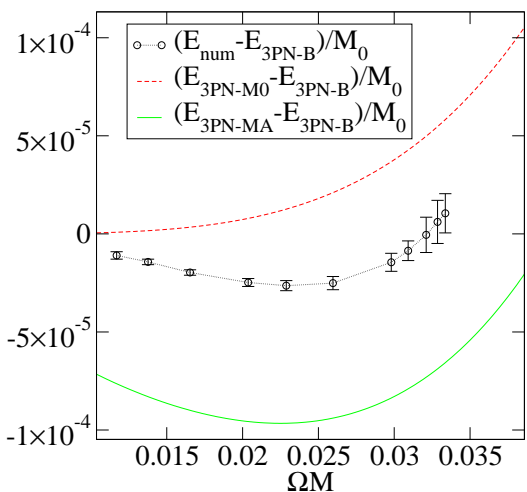}
\caption{\label{fig:DiffE} Difference between our results for the binding energy
of a sequence of quasiequilibrium equal-masses binaries, and the 3PN predictions
from model 3PN-B. The errors represented here come from the difference between the 
ADM and the Komar
energies, except for the 3 closest binaries, for which the numerical error can no
longer be neglected.\\
We also represent the influence of tidal effects (from model 3PN-M0) and eccentricity
(model 3PN-MA). Any binary with an eccentricity $e=0.01$, initially closer to its apocenter
than to its pericenter, should have a binding energy between the results from models 3PN-M0
and 3PN-MA. Model 3PN-MP, representing an eccentric binary at its pericenter, is not plotted
here, but predicts even higher energies than model 3PN-M0.}
\end{figure}

In Fig.\ \ref{fig:Eb}, we show results for the binding energy for various binary separations, where both numerical simulations seem to be
in good agreement. 
For our results, the precision reached is good enough to measure deviations from the 3PN predictions neglecting
eccentricity. We observe 
differences of order $10^{-5}$ for configurations where our expected precision is about an order of magnitude better. 
In Fig.\ \ref{fig:DiffE}, we show the deviations from the simplest 3PN model (3PN-B) over a large range 
of separations. The behavior
at small separation is not resolved well enough to note anything other than the divergence of the numerical and 3PN
predictions when the star reaches its disruption point. But for most of the sequence, we observe that the numerical 
results are clearly below the 3PN predictions, the difference between the two results decreasing at the largest separations.

It is also easy to see that tidal effects cannot explain these results. They contribute at the same order of magnitude, but
tend to \emph{increase} the energy of the system. However, Fig.\ \ref{fig:DiffE} shows that our results are still compatible with the 
3PN predictions if we include the influence of eccentricity. Indeed, its effects can decrease the energy of the system if we are closer
to the apocenter than the pericenter --- and, in fact, we know from short evolutions that this is the case for our initial data (see Table
\ref{tab:lowecc}).

\begin{figure}
\includegraphics[scale=0.48]{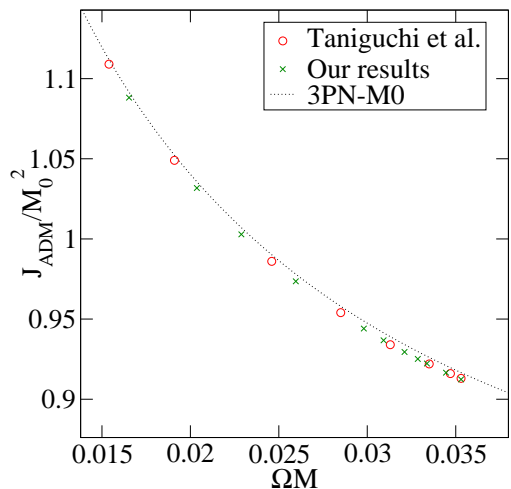}
\caption{\label{fig:J} Angular momentum of equal-mass binaries for initial data
generated by our solver, initial data from Taniguchi et al. ~\cite{Taniguchi2008}, and 3PN
predictions. We see that both numerical results are in very good agreement,
even when they begin to diverge from the 3PN models of circular orbits. }
\end{figure}

A similar comparison can be made using the total angular momentum $J_{\rm ADM}$, as shown in Fig.\ \ref{fig:J}.  
The agreement between both numerical calculations is clearly visible, even in the regime where they deviate from the
3PN models of circular orbits. This should not be surprising, as both sets of numerical results use essentially the same formulation of the problem. 
As was the case for the energy, results for $J_{\rm ADM}$ can only be reconciled
with the 3PN predictions if we assume a small eccentricity and an initial state closer to the apocenter than to the pericenter.

Overall, these results show that our precision is good enough to resolve deviations from the point-mass, circular orbit 3PN predictions.
We have thus the potential to study the main effects contributing to these deviations in irrotational BH-NS binaries: tidal effects in the neutron star,
influence of the eccentricity of the orbit,
and spurious gravitational effects due to the inconsistency of the quasiequilibrium formulation.

\subsection{Spinning black holes}
\label{sec:SpinBH}
For nonspinning or slowly spinning BHs, the conformally flat metric we have used until now performs extremely well. However, if we want
to generate rotating BHs, being able to use a different conformal metric is critical.
The natural choice for such BHs would be a Kerr-Schild conformal metric. Unfortunately, early results from Taniguchi et al.\ ~\cite{Taniguchi2006}
have shown that this leads to strong deviations from quasiequilibrium: the difference between the Komar and ADM energies in \ ~\cite{Taniguchi2006}
is of the order of the binding energy. Here, we describe the use of a modification of the Kerr-Schild metric, 
already applied to BBH by Lovelace et al.\ ~\cite{Lovelace2008}. 

We define $\gamma_{ij}^{KS}(a_{\rm BH},{\bf v}_{\rm BH})$ and $K^{KS}(a_{\rm BH},{\bf v}_{\rm BH})$ as the 3-metric and trace of the extrinsic curvature
of a black hole with spin parameter $a_{\rm BH}$ and boost velocity ${\bf v}_{\rm BH}$, written in Kerr-Schild coordinates. Then, we choose the
free parameters of the XCTS equations as follows:
\bea
\tilde \gamma_{ij}&=&\delta_{ij}+[\gamma_{ij}^{KS}(a_{\rm BH},{\bf v}_{\rm BH})-\delta_{ij}] e^{-(r_1/w)^4},\\
K&=&K^{KS}(a_{\rm BH},{\bf v}_{\rm BH}) e^{-(r_1/w)^4},\\
{\bf v}_{\rm BH}&=&{\bf \Omega} \times {\bf c}_{\rm BH}
\eea
where $r_1$ is the coordinate distance to the center of the BH ${\bf c}_{\rm BH}$, and the width $w$ is chosen as half
the coordinate distance between the two compact objects. This choice ensures that close to the BH, the metric is nearly $\gamma_{ij}^{KS}$, 
while away from the hole, we recover conformal flatness and maximal slicing. The introduction of the exponential  damping $e^{-(r_1/w)^4}$ 
is the most important difference between the choices of conformal metric and extrinsic curvature in\ ~\cite{Lovelace2008} 
and\ ~\cite{Taniguchi2006}. That change is indeed necessary to avoid large deviations from equilibrium.
 
To take advantage of the similarities between this initial configuration and a Kerr black hole, we also change the boundary condition
imposed on the lapse. If the lapse of an isolated Kerr-Schild BH is $\alpha^{KS}(a_{\rm BH},{\bf v}_{\rm BH})$, our boundary
condition on the excision surface will be
\beq
\label{eq:BCKS}
\alpha = \alpha^{KS}(a_{\rm BH},{\bf v}_{\rm BH}) e^{-(r_1/w)^4}
\eeq
instead of (\ref{eq:BCgauge}). To get as close as we can to a Kerr-Schild BH, we modify the shape of the excision surface. 
The subdomain containing the apparent horizon is now a spherical shell in coordinates $(r_K,\theta,\phi)$. The Kerr radius $r_K$ is
defined as the largest positive root of the equation 
$r_K^4-r_K^2(r^2-a^2)-({\bf a \cdot r})^2=0$, where $r$ is the coordinate distance to the center of the BH, and ${\bf a}={\bf J}/M^{\rm ADM}_{\rm BH}$
is the spin parameter. The excision surface is then the oblate surface $r_K = {\rm constant}$, and we choose the constant so that 
$M^{\rm ADM}_{\rm BH}=1$.

Once these choices have been made, no further modifications of our numerical methods are required. We test the performance of these new data sets
on two types of BH-NS binaries. First, we consider configurations with a nonspinning BH, which allows direct comparison with the conformally
flat initial data. Then, we move to BHs with a spin $J_{\rm BH}=0.5 (M^{\rm ADM}_{\rm BH})^2$, with the direction of $J_{\rm BH}$ opposite 
to the orbital angular momentum, and verify that comparable results can
be obtained. Tables \ref{tab:NF} and \ref{tab:Spin} summarize the properties of the resulting binaries.
Different spins aligned with the rotation axis can be obtained using the same method. We tested our procedure up to spins of 0.9, and note that, as for BBH initial data\ ~\cite{Lovelace2008},
the deviations from quasiequilibrium tend to increase with the spin of the BH (the difference between Komar and ADM mass reaches about 10\% of the binding energy for a spin of 0.9).
The choices of the conformal metric and the lapse boundary condition seem 
to have a major influence on the amplitude of these deviations. Better choices will probably help reduce the deviations observed for rapidly rotating BHs.

\begin{table}
\caption{Same as Table \ref{tab:MR1}, but for BH-NS binaries built with a modified Kerr-Schild
conformal metric, as described in Sec.\ \ref{sec:SpinBH}. The spin of the BH is still 0. }
\label{tab:NF}
$\begin{array}{|c|c|c|c|c|}
\hline
\frac{d}{M_0} & \Omega M_0 & \frac{E_b}{M_0} & \frac{J}{M_0^2} & \frac{|E_{\rm ADM}-M_K|}{M_0}\\
\hline
18.489 & 0.01171 & -6.15 \times 10^{-3} & 1.195 & 8.4\times 10^{-6} \\
16.990 & 0.01321 & -6.64 \times 10^{-3} & 1.155 & 8.8\times 10^{-6} \\
15.490 & 0.01507 & -7.20 \times 10^{-3} & 1.116 & 9.3\times 10^{-6} \\
13.991 & 0.01741 & -7.87 \times 10^{-3} & 1.074 & 1.0\times 10^{-5} \\
12.491 & 0.02042 & -8.67 \times 10^{-3} & 1.032 & 1.1\times 10^{-5} \\
11.492 & 0.02295 & -9.30 \times 10^{-3} & 1.003 & 1.1\times 10^{-5} \\
10.493 & 0.02605 & -1.00 \times 10^{-2} & 0.974 & 1.3\times 10^{-5} \\
\hline
\end{array}$
\end{table}

\begin{table}
\caption{Same as Table \ref{tab:NF}, but the BH now has a spin $J_{\rm BH}=-0.5$. }
\label{tab:Spin}
$\begin{array}{|c|c|c|c|c|}
\hline
\frac{d}{M_0} & \Omega M_0 & \frac{E_b}{M_0} & \frac{J}{M_0^2} & \frac{|E_{\rm ADM}-M_K|}{M_0}\\
\hline
18.368 & 0.01182 & -6.03 \times 10^{-3} & 1.081 & 2.9\times 10^{-5} \\
16.881 & 0.01335 & -6.50 \times 10^{-3} & 1.043 & 3.4\times 10^{-5} \\
15.395 & 0.01523 & -7.04 \times 10^{-3} & 1.004 & 4.0\times 10^{-5} \\
13.908 & 0.01759 & -7.68 \times 10^{-3} & 0.964 & 4.7\times 10^{-5} \\
12.422 & 0.02065 & -8.43 \times 10^{-3} & 0.924 & 5.6\times 10^{-5} \\
11.431 & 0.02321 & -9.01 \times 10^{-3} & 0.896 & 6.2\times 10^{-5} \\
10.441 & 0.02636 & -9.67 \times 10^{-3} & 0.869 & 6.8\times 10^{-5} \\
\hline
\end{array}$
\end{table}

We first note that, for equivalent resolutions, the new configurations are less precise by typically an order of magnitude. Also, as we want the width $w$
to be large compared to the radius of the apparent horizon, we should avoid close binaries. Deviations from quasiequilibrium 
are also significantly larger, but not nearly as much as in ~\cite{Taniguchi2006}, where an unmodified Kerr-Schild background was used. In ~\cite{Taniguchi2006},
the difference between Komar and ADM energies was of the order of the binding energy, while here it is only about 0.15\% of that value. Direct
comparison between our results for a flat conformal metric and Table \ref{tab:NF} also shows that both sets of initial configurations are in agreement. 

As long as the BH is not rotating, the new conformal metric does not lead to any noticeable advantage over the conformally flat background --- though the
initial burst of gravitational radiation might end up being smaller. For rotating BHs, however, a conformally flat metric is no longer appropriate, while
a modified Kerr-Schild metric allows us to solve the initial data problem. Deviations from quasiequilibrium will increase once more, but for a BH spin
$J_{\rm BH}=-0.5$, we can still solve for the binding energy within a fraction of a percent. The norm of the constraints is also below $5 \times 10^{-6}$ for our closest
binaries, making these initial configurations
perfectly suitable for future evolutions.

Our ability to reach high accuracy at a relatively low resolution is particularly important for the construction of these spinning configurations.
Indeed, the slower convergence rate makes it significantly harder to obtain useful initial data. Moreover, as the geometry around each compact object become less
and less spherical, being able to easily adapt our numerical grid becomes even more necessary. These first results show that the construction of spinning BHs is perfectly
possible without much modification of our basic formalism --- and improvements in the choice of the conformal metric and/or the excision boundary conditions 
might further improve the quality of such initial configurations.

\subsection{Low-eccentricity binaries}
\label{sec:lowecc}

\begin{table}
\caption{Orbital parameters of three irrotational BH-NS binaries, after 0, 1, and 2 steps of the iterative procedure
designed to reduce the eccentricity of their orbits. The initial radial velocity of an observer comoving with the NS is
$v_r=\dot{a}_0 d_0/2$, the eccentricity is measured from the parameters of the fit (\ref{eq:ellfit}) according to $e=B/\omega d_0$, 
and the orbital phase $\phi$ is $0$ at pericenter and $\pi$ at apocenter. }
\label{tab:lowecc}
$\begin{array}{|c|c|c|c|c|c|}
\hline
 & v_r & \Omega M_0 & {\it e} & \phi/\pi & \frac{|E_{\rm ADM}-M_K|}{M_0}\\
\hline
{\rm Step 0} & 0 & 0.02157 & 1.0 \times 10^{-2} & 0.68 & 2.3 \times 10^{-6}\\
{\rm Step 1} & -9.36(-4) & 0.02161 & 4.4 \times 10^{-3} & 1.18 & 2.8 \times 10^{-4}\\
{\rm Step 2} & -7.20(-4) & 0.02165 & 6.5 \times 10^{-4} & 1.59 & 2.9 \times 10^{-4}\\
\hline
\end{array}$
\end{table} 

\begin{figure}
\includegraphics[scale=0.48]{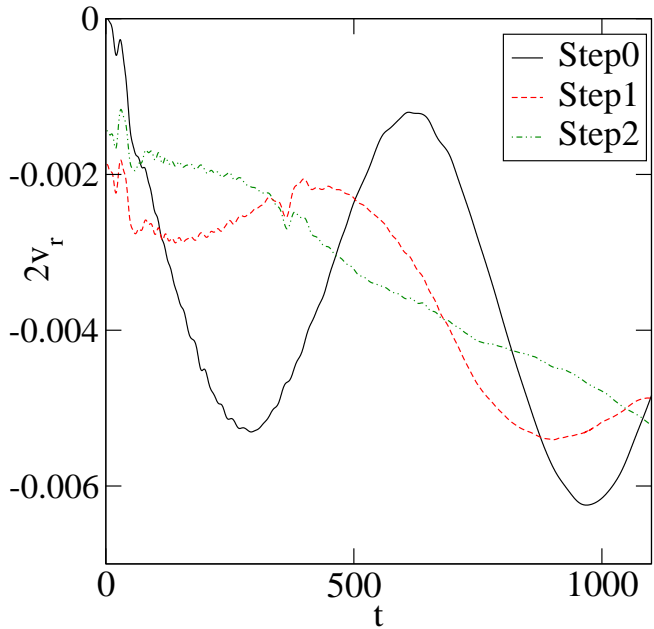}
\caption{\label{fig:lowellda} Evolution of equal-mass binaries after 0, 1 and 2 steps of our iterative method reducing
the eccentricity. We plot the time derivative of the coordinate separation between the BH and the NS, $2v_r=\dot{d}$. }
\end{figure}

The initial configurations discussed in this paper correspond to binaries only a few orbits away from merger. Such systems are expected to have nearly
circular orbits as, because of gravitational wave emission, the eccentricity decreases as a power law of the distance between the objects ~\cite{Peters1964}.
The influence of the eccentricity on observable quantities such as the gravitational waveform can be significant.
For instance, it is one of the dominant effects limiting the comparison between high-accuracy BBH evolutions and post-Newtonian expansions presented in
~\cite{Boyle2007}, even though the initial eccentricity of their binary is lower than $6\times 10^{-5}$.

Evolutions of BH-NS systems are far from being as precise. But the force balance condition we used until now leaves the binaries with eccentricities 
of order 0.01 --- enough to be noticeable in evolutions. We thus want to decrease the eccentricity of the initial data so that its influence
on the orbit is at most of the order of the precision of the evolution code. 

Here, we show that the iterative method already used to reduce 
the eccentricity of BBH ~\cite{Pfeiffer2008} can be applied successfully to BH-NS binaries. For all evolutions described in this section,
we used the mixed finite difference-spectral code described in Duez et al. ~\cite{Duez2007}.

The eccentricity and orbital phase of our binaries are determined by the choice of orbital angular velocity $\Omega$ and infall velocity $\dot{a}_0 {\bf r}$.
Until now, we have been determining $\Omega$ through Eq. (\ref{eq:forcebalance}), choosing $\dot{a}_0=0$. Now, we will use such configurations
as a first approximation to the low-eccentricity solution, and try to determine from its evolution better values of $\Omega$ and $\dot{a}_0$.

To do so, we record the coordinate separation between the center of the compact objects, $d$, and fit its time derivative by the formula
\beq
\label{eq:ellfit}
\dot{d}=A_0+A_1 t + B \sin{(\omega t + \phi)},
\eeq
where the parameters $A_0$, $A_1$, $B$, $\omega$, and $\phi$ are all determined by the fit. For a Keplerian orbit, we would have $A_0=A_1=0$, and an
eccentricity $e=B/\omega d_0$, where $d_0=d(t=0)$. We use this definition of $e$ as an approximation of the eccentricity of the system.  As in ~\cite{Pfeiffer2008}, 
we then choose the corrections to $\Omega$ and $\dot{a}_0$ so that a Keplerian orbit with the same parameters $d$, $\omega$, $\phi$,
and $B$ would become circular:
\bea
\delta \dot{a}_0 & = & -\frac{B \sin{\phi}}{d_0},\\
\delta \Omega & = & - \frac{B \omega \cos{\phi}}{2 d_0 \Omega_0}.
\eea
For the fit (\ref{eq:ellfit}) to be accurate, we need to evolve the binaries for at least one and a half orbits. Furthermore, as the initial spurious burst
of gravitational radiation in the data disturbs the early motion of the binary, we also exclude points at $t<100 M$ from the fit. 

As a first example, we consider a binary at initial coordinate separation $d/M_0=12.0$, and evolve it using the fully relativistic numerical code described in
~\cite{Duez2007}. From this evolution, we determine that the eccentricity of the initial data constructed by requiring force balance (\ref{eq:forcebalance}) and
$\dot{a}_0 {\bf r}=0$ is of order ${\it e}=0.01$. We then go twice through the iterative method we just described. The orbital parameters of the three binaries we
evolved are listed in Table \ref{tab:lowecc} while in Fig.\ \ref{fig:lowellda}, we show the time derivative of the coordinate separation, $\dot{d}$. Two iterations
reduce the eccentricity by about an order of magnitude. Decreasing the eccentricity further would demand evolutions at a higher resolution, increasing the computational
cost, but does not in principle involve any new difficulties. We also find that the difference between ADM energy and Komar mass increases by about 
2 orders of magnitude during eccentricity removal (see Table \ref{tab:lowecc}).

\section{Discussion}

In this paper, we presented a new method for the construction of initial data for BH-NS binaries, based on the multidomain
spectral elliptic solver {\sc spells} ~\cite{Pfeiffer2003b}. The flexibility of the multidomain spectral methods allows the use
of a numerical grid adapted to the geometry of the system. We showed that this allows us to build high-accuracy initial
data while keeping the number of grid points relatively low.

Using the extended conformal thin sandwich formalism and fixing the initial state of the system through quasiequilibrium conditions,
we obtained initial data whose precision is limited only by the small deviations from an exact equilibrium. 
As an example, we showed convergence tests for a sequence of equal-mass, irrotational
BH-NS binaries, verifying the exponential convergence of our solver. Corotational and unequal-mass systems lead to similar results.

We also showed that with such accuracy we can resolve deviations from the point mass, circular orbit 3PN predictions, and observe
the influence of tidal distortion and eccentricity.

Abandoning the assumption of conformal flatness, we generalized the method to construct binaries with a spinning black hole.
Previously, initial data with a Kerr-Schild conformal metric was shown to be significantly inferior to conformally flat configurations
~\cite{Taniguchi2007}. Here, we showed that using a Kerr-Schild metric cut off at large distances from the BH allows reasonable precision
to be reached--- as in the case of BBH ~\cite{Lovelace2007}. We verified that with such
a conformal metric we could construct a binary whose BH has a spin $J_{\rm BH}=-0.5$ perpendicular to the orbital plane.

Finally, we adapted a method designed for BBH ~\cite{Pfeiffer2008}, and demonstrated our ability
to significantly decrease the eccentricity of the binary initial data.

\begin{acknowledgments}
It is a pleasure to acknowledge useful discussions with Matthew Duez, Eanna Flanagan, Jan Hesthaven, Francois Limousin, Geoffrey Lovelace, 
Robert Owen, Mark Scheel and Manuel Tiglio. In particular, we would like to thank Matthew Duez for his help in evolving low-eccentricity
binaries, Jan Hesthaven for his contribution to the improvement of the preconditioner in {\sc spells}, Geoffrey Lovelace for his advice on the
construction of spinning black holes, and Robert Owen for the measurement of these spins. This work was
supported in part by grants from the Sherman Fairchild Foundation to
Caltech and Cornell, and from the Brinson Foundation to Caltech; by
NSF Grants No. PHY-0652952, No. DMS-0553677,
and No. PHY-0652929, and NASA Grant No. NNG05GG51G at Cornell; and by
NSF Grants No. PHY-0601459, No. PHY-0652995 and NASA Grant
No. NNG05GG52G at Caltech. 
\end{acknowledgments}

\bibliography{References/References}

\end{document}